# Femtosecond laser-controlled production of functional titanium niobate layers from $TiO_2/Nb_2O_5$ powders

Gonzalo Gomez-Munoz[1*], Aarón Calvo-Villoslada[2], Inmaculada Álvarez-Serrano[3], María Luisa López[3], Paloma Fernandez[2], Javier Solis[1], Belén Sotillo[2**].

[1]Laser Processing Group (LPG), Instituto de Óptica "Daza de Valdés" (IO-CSIC), CSIC, Serrano 121, 28006, Madrid, Spain.
[2]Departamento de Física de Materiales, Facultad de Ciencias Físicas, UCM, Plaza de Ciencias 1, 28040, Madrid, Spain.
[3]Departamento de Química Inorgánica, Facultad de Ciencias Químicas, UCM, Plaza de Ciencias 2, 28040, Madrid, Spain.

[*]gonzalo.gomez@io.cfmac.csic.es
[**]bsotillo@ucm.es

**Abstract:**

Conventional synthesis methods of titanium niobates (TNO) require prolongated high-temperature treatments. To overcome this limitation, in this work, we have developed a confined powder-bed femtosecond (fs) laser-based technique to sinter TNO films starting from a mixture of $TiO_2$ and $Nb_2O_5$ powders, avoiding furnace-based approaches. The influence of laser-processing conditions on film formation, phase composition, and microstructure is systematically investigated, demonstrating that different TNO phases and morphologies can be obtained by controlling the laser pulse energy. A comprehensive characterization of the processed films has been performed, providing insights into the mechanisms governing their formation, which appear to share key features with laser-induced backward transfer (LIBT) processes. Our findings establish the relationship between laser-processing conditions and TNO film formation, highlighting the potential of confined powder-bed fs-laser processing as an alternative route for manufacturing functional oxide layers. Tentative applications of these layers are finally discussed.

## 1. Introduction:

Titanium niobates (TNOs) are a large family of ternary oxides, based on titanium (Ti) and Niobium (Nb), in which several polymorphs show a rich variety of properties that exhibit a potential interest for their implementation into technological applications. In particular, the use of some of these polymorphs ($TiNb_2O_7$, $Ti_2Nb_{10}O_{29}$…) in the development of environmental technologies, such as lithium (Li) ion batteries (LIBs) [1] or catalytic devices for pollutants degradation and hydrogen production [2], has lately attracted a lot of attention. Despite the TNOs good characteristics, there are key limitations to their industrial use, related to the high energy and time costs involved in producing the material with an adequate control of its structure and composition.

TNOs (with general formula $TiNb_xO_{2+5x/2}$) are usually understood from the perspective of the $TiO_2$-$Nb_2O_5$ system [3], and share with $Nb_2O_5$ the Wadsley-Roth crystalline structure [4], derived from $ReO_3$ octahedral structure [5]. $TiNb_2O_7$ (x = 2), which crystallizes in the monoclinic system, is composed of 3 × 3 $ReO_3$-type blocks of corner-sharing [$MO_6$] octahedra (M=Nb, Ti); adjacent blocks are linked through edge-sharing octahedra, forming crystallographic shear planes (see **Figure 1**). The Ti and Nb cations are randomly distributed over the octahedral metal sites throughout the structure [6]. This stoichiometry is the most common and stable one and has shown a noticeable interest in the energy storage field because of its great potential for Li accommodation, triggered by its open Wadsley-Roth structure, which reduces the structural stress caused by the insertion of Li in the TNO [7]. This fact, together with the high cyclability achieved thanks to the stability conferred by the nature of the bonds between the different elements (covalent) and by edge-sharing, make this and other TNO polymorphs [8] interesting candidates for the development of LIBs [9,10].

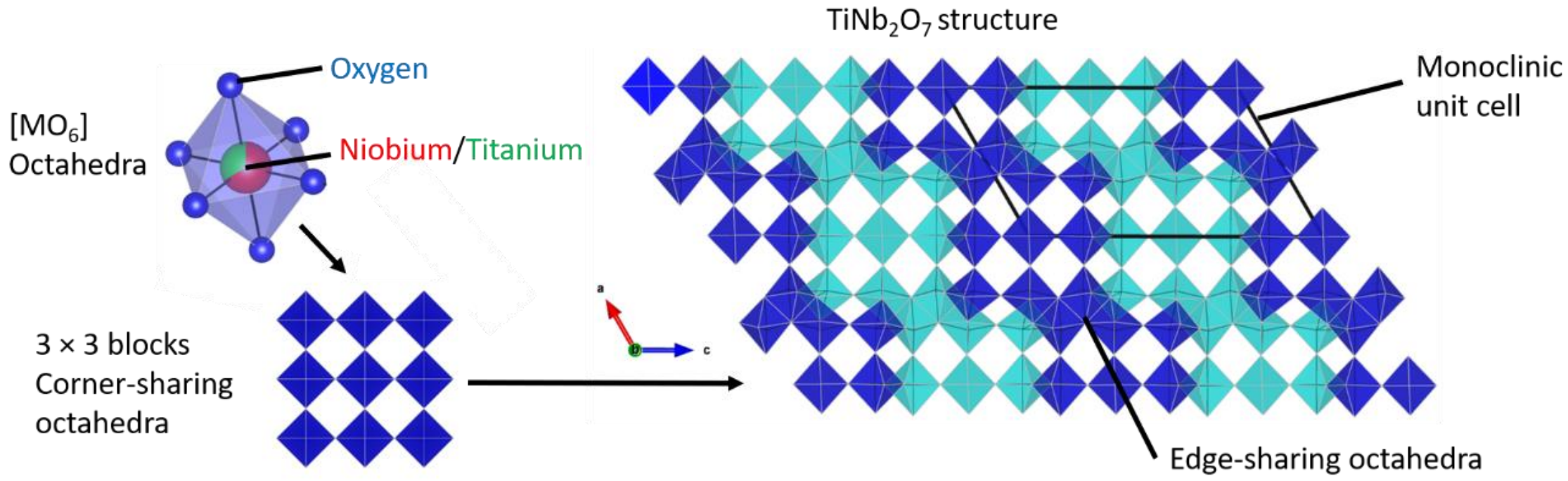

**Figure 1.** Crystal structure of monoclinic $TiNb_2O_7$, shown from the basic octahedra building unit to the 3 × 3 $ReO_3$-type blocks and the complete crystal structure. Octahedra shown in different shades of blue represent $ReO_3$-type blocks located in different planes perpendicular to b-axis. The structures were plotted using VESTA software [11].

The fabrication of TNOs relies on energy- and time-consuming processes. The usual TNO production methods, such as solid-state [7,12] and solvothermal [13,14] synthesis, require high temperatures (up to 1900 ºC) to achieve the material crystallization [3]. In addition, several $TiO_2$-$Nb_2O_5$ Wadsley-Roth phases, including $TiNb_2O_7$ and $Ti_2Nb_{10}O_{29}$, have been reported to be entropy-stabilized at elevated temperatures, further highlighting the importance of the thermal history during their synthesis [15]. Alternative methods [1] still need a high temperature annealing or calcination process (either of the precursors or of the synthesized material) to achieve the targeted phase of the TNO. For this reason, identifying new TNO synthesis routes based on techniques that are less demanding in terms of time and energy consumption is essential for the development of clean and sustainable technology.

In this work, we propose a femtosecond (fs) laser-based technique to controllably produce functional TNO films. Fs-lasers are a powerful tool that has been used since its invention to modify any kind of materials and their properties [16], allowing or enhancing their performance in multiple applications. Fs-lasers pulses reach peak intensities up to 10 $TW \cdot cm^{-2}$ and, when focused, interact with the target within a micrometric volume. Therefore, they can be used to drive non-linear and multiphoton absorption processes. These features allow fs-lasers to deposit a large amount of energy in a localized way, making ultrafast processing a useful tool for melting and phase change induction [17]. Thus, fs-laser processing appears as an attractive alternative to conventional solid-state and solvothermal methods, introducing the possibility of sintering ceramic materials in very short times while avoiding the use of high-temperature furnaces or autoclaves.

In a previous work [18], we demonstrated that it was feasible to sinter layers of high temperature monoclinic $Nb_2O_5$ (H-$Nb_2O_5$) by fs-laser powder processing of $Nb_2O_5$ pellets (discs of compacted powders of 2 mm in thickness and 7 mm in diameter). The laser-sintered layers presented improved electrical behaviour when compared to the usual oxide, showing that the fs-laser pulses triggered the reduction of Nb in the oxide lattice. However, this approach presented some limitations that motivated the development of the processing strategy here proposed. On the one hand, during the preparation of the pellet, a thermal treatment (700 ºC, 2 h) was needed to avoid the crumbling of the pellet upon the laser irradiation, partially limiting the advantages associated with direct laser processing. On the other hand, it was impossible to separate the sintered ceramic film from the rest of the pellet without breaking the layer, which restricted the subsequent handling and use of the processed layers. In addition, it has been reported that introducing additional species in pure $Nb_2O_5$ (Ti in our case) can modify the oxygen-defect chemistry and electrical behaviour of the oxide, while potentially improving its performance in functional devices [19,20].

Building on our previous results reported in [18], here we have developed an advanced fs-laser processing strategy in which the target powder, consisting of a mixture of $TiO_2$ and $Nb_2O_5$, is directly irradiated, avoiding the high-temperature powder preparation steps and facilitating the recovery of the processed layers. Depending on the energy of the laser pulses used, three different scenarios may be found, leading to the formation of TNO layers with different composition and crystalline structure. In this work, a direct relationship between the laser irradiation conditions and the functional properties of the produced layers is established.

## 2. Experimental section:

### 2.1 Powders preparation

The starting material consists of a mixture of two different precursors: powders of $Nb_2O_5$ (Alfa Aesar Puratronic®, 99.9985% purity, particle size < 10 μm) and $TiO_2$ (Sigma-Aldrich, 99.99% purity, particle size < 45 μm). These powders are weighted with a precision balance to ensure an atomic ratio between Ti and Nb which points to the most stable TNO stoichiometry: Ti (at%):Nb (at%) ratio of 1:2, corresponding to $TiNb_2O_7$, which is going to be the target composition in our experiments. After weighting, the powder was ball milled (Retsch S100) for 5 hours at 300 rpm with 20 mm agate balls.

### 2.2 Design of irradiation targets

Preliminary tests based on the pellet-processing strategy previously reported in [18] were firstly performed; the corresponding experimental details and results are provided in the Supplementary Material File (**Figure SM1**). Based on these preliminary tests, a confined powder configuration was adopted for the experiments described here. One gram of powder mixture was placed in an aluminium bed (see **Figure 2**) covering an area of 2.5 × 2.5 $cm^2$ and a thickness of ~1 mm. The mixture of powders is then compressed between the aluminium bed and a Fused Silica Glass (FSG, Spi Supplies®, 1 mm thick Corning® 7980 Synthetic Flame Fused Silica), as shown in **Figure 2(A)**. Two metallic clamps allow to hold the FSG, which acts as a barrier to prevent the powder fall (**Figure 2(B)**). After laser irradiation, the sintered material shows a different coloration than the pristine powder, as it is shown in **Figure 2(C)**.

### 2.3 Femtosecond-laser irradiation

The fs-laser irradiation setup is shown in **Figure 2(D)**. The laser source is an Yb-doped fs-pulsed fiber laser (Satsuma HP, Amplitude Systemes) operating at a central wavelength $\lambda$ = 1030 nm, with a pulse duration $\tau$ = 350 fs and a pulse repetition rate $f_{rep}$ = 500 kHz. The beam is steered through a scanning system composed by a galvanometric mirror scanner (ScanCube 14, Scanlab) and a F-Theta lens (f = 100 mm), allowing to scan the laser focus over a stationary sample surface; the combination of these two elements provides a flat field focusing over an area of 7 × 7 $cm^2$. The laser beam polarization was linear and controlled using a $\lambda/2$ waveplate (HLP) to be vertical on the sample, parallel to the scanning direction (see **Figure SM2**). A 5.5 mm pinhole has been used to enlarge the focus depth of the lasing system. The beam spot diameter was $2\omega_0$ = 37 μm ($1/e^2$ intensity), previously calculated on a Silicon piece according to the method proposed by Liu in [21]. The calculated focal depth was approximately 1340 μm.

After the optimization of the focus position (described in **Section 3.1**), initial energy-screening experiments were performed in 2 × 2 $mm^2$ areas, modifying the pulse energy (E) in each irradiation in the range of 1 - 8 μJ, in steps of 1 μJ. All the pulse energies given in this article are calculated before considering the transmittance of the lens used for focusing the beam (which is 83% transparent at the laser wavelength) to give round values and to facilitate the comparison. Each area was produced using a single raster pass composed of unidirectional parallel scanning lines. Following our previous experiments, the scanning speed was first fixed at v = 0.10 m/s and the separation between consecutive scanning lines was fixed at d = 4 μm [18]. These experimental conditions allow the selection of three representative pulse energy conditions (E = 1, 4 and 6 μJ), which were subsequently used to perform detailed characterization of the material produced. Additional 6 × 6 $mm^2$ areas were produced at each of these pulse energies to enable further measurements, as described in **Section 2.4**.

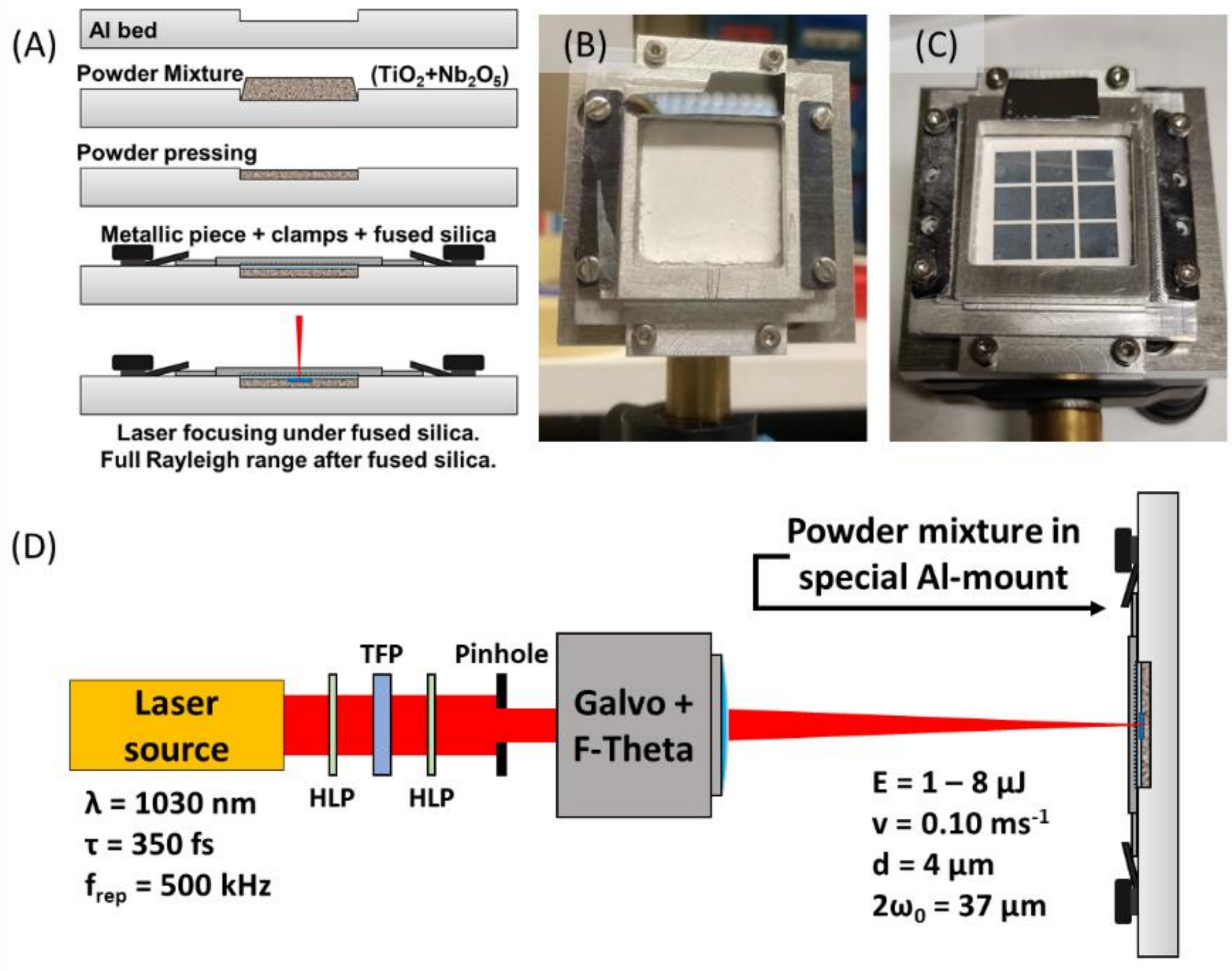

**Figure 2.** (A) Powder preparation scheme. (B) Appearance of the mixture of powders as prepared for the irradiation and (C) just after the fs-laser process (seen in both cases through the FSG). (D) Irradiation layout showing the femtosecond-laser system used. In the scheme, the irradiation parameters are indicated. Also, the relative positions of the HLP, thin film polarizer system (TFP) and pinhole used in the setup are shown.

### 2.4 Recovery and characterization of the laser-sintered material

Depending on the pulse energy E, different scenarios can be found after removing the FSG from the aluminium mount: transference (**Figure 3(A)**) or non-transference (**Figure 3(B)**) of the sintered material to the FSG. The scenario determines the post-laser processing characterization because of the presence or absence of a rigid substrate that sustains the processed areas (**Figure 3(C)**). For the transference scenario, the samples remain attached to the FSG surface after removing the glass from the holder, so the excess powder is removed using a nitrogen spray gun, followed by ultrasonication in distilled water before characterization. For the non-transference scenario, the processed areas remain on the powder bed and the films processed are continuous but fragile, so they are recovered using a spatula. The excess powder is removed with a nitrogen spray gun, and the samples are then placed on a copper sample holder.

To analyse the morphology and appearance of the samples, two different microscopy techniques have been applied: Optical Microscopy (OM) and Scanning Electron Microscopy (SEM) in secondary electron (SE) and/or backscattering electron (BSE) modes. The OM used was a Leica MSV266 under white illumination with a Leica DFC295 camera mounted. The SE-SEM images were taken with a FEI Inspect S working at 15 kV and the BSE-SEM images with a Hitachi TM3000 Tabletop SEM working at 20 kV. This last SEM was also used to study the composition of all the samples by Dispersive X-Ray microanalysis (EDX), using a Brucker AXS Quantax system.

The structural characterization has been performed by the combination of X-Ray Diffraction (XRD) and Raman Spectroscopy (RS). XRD measurements were carried out in goniometer configuration using Cu Kα radiation and a 2θ step size of 0.05°. Measurements on the transferred layers were performed using a Bruker D8 Advance A25 diffractometer in standard Bragg-Brentano θ-2θ geometry with a fixed divergence slit, whereas measurements on the non-transferred layers were conducted in constant radiated length mode using a Malvern Panalytical Empyrean diffractometer, with a spotsize of 4.5 × 2.5 $mm^2$. The corresponding diffraction data were converted and analysed using X'Pert Highscore software.

Micro-RS measurements were carried out in a Horiba JobinYvon LABRAM-HR confocal microscope, using the 632.8 nm line of a He-Ne laser. The laser was focused onto the layers with a 100× Olympus objective (0.9 NA), which was also used to collect the scattered light (backscattering configuration), and the signal was analysed with a 600 lines/mm grating and an air-cooled charge coupled device camera (CCD). The spot size of the laser beam for the selected objective is ~ 1 μm. The software used to perform the RS analysis was the Labspec 5.0 software of the confocal microscope. In the case of the transferred layers and thanks to the hardness of the samples, cross-sectional SEM (X-SEM) and micro-RS (X-RS) were performed to better understand the sintering process.

To evaluate the functional properties of the transferred layers produced at 4 and 6 μJ were characterized electrically, whereas electrochemical measurements were performed on the non-transferred samples obtained at 1 μJ. This is because non-transferred samples are free-standing when removed from the powder bed and cannot be equally characterized than those attached to a FSG, as it will be discussed in **Section 3.5**.

The electrical properties of transferred layers were evaluated by four-point-probe measurements, performed with a 4-point-probe Ossila® device. The probe spacing was 1.27 mm, and the target current for all the measurements was 8 μA, while the applied voltage was increased in steps of 100 mV until reaching a maximum value of 10 V. Each direct sheet resistance ($R_\square$) measurement was repeated 100 times in each laser-sintered area of interest to ensure the study's statistical accuracy. Then, the bulk resistivity (ρ), which also depends on the layer thickness, is assessed as $\rho = R_\square \cdot$thickness.

For electrochemical characterization, the 1 μJ material was used to prepare composite electrodes containing 70 wt.% active material, 20 wt.% carbon black (Ensaco, conductive additive), and 10 wt.% sodium alginate (Sigma-Aldrich). The components were dispersed in distilled water until a homogeneous slurry was obtained and subsequently coated onto a copper foil. Half-cells were assembled in Swagelok-type cells using a Whatman GF/D borosilicate glass-fibre separator, 1 M $LiPF_6$ in EC:DMC (1:1 by weight) as the electrolyte, and metallic lithium as the counter/reference electrode. Cell assembly was carried out in an argon-filled glovebox.

Electrochemical measurements were performed using a BioLogic 815 system. Cyclic voltammetry (CV) was conducted between 1 and 3 V, while galvanostatic cycling with potential limitation (GCPL) was performed at different charge/discharge rates (C-rates). These rates describe the speed at which a battery stores or releases energy and is defined relative to its theoretical capacity. For instance, a rate of 1C corresponds to a full charge or discharge in 1 h, whereas 2C and C/2 correspond to 30 min and 2 h, respectively.

Potentiostatic electrochemical impedance spectroscopy (PEIS) measurements were carried out using a 5 mV AC perturbation over the frequency range from 10 kHz to 1 Hz. Within this frequency window, the response is mainly governed by charge-transfer processes and the onset of diffusion-controlled phenomena, whereas the high-frequency processes associated with the bulk electrolyte resistance and the low-frequency capacitive contributions related to long-term charge accumulation remain only partially accessible. Consequently, the impedance response was interpreted considering the processes that are expected to dominate within the explored frequency interval [22]. The PEIS spectra were fitted using equivalent-circuit models by complex nonlinear least-squares fitting with modulus weighting. The circuit was adapted after cycling to account for the additional low-frequency transport contribution.

## 3. Results and discussion:

In this section we will cover the experimental results achieved for different group of laser processing conditions.

### 3.1 Determination of the processing window.

Preliminary irradiations on pellets (see Supplementary Material File) showed that promoting Ti incorporation required severe irradiation conditions, which led to material ejection, while the processed layers were difficult to separate from the pellet (see **Figure SM1**). These limitations motivated the development of the confined powder configuration setup shown in **Figure 2**.

The confined configuration was first evaluated by varying the position of the focal volume relative to the FSG-powder interface at E = 4 µJ and v = 0.10 m/s (**Figure 3(D)**). The goal was to maximize the volume of powder exposed to the high-intensity region (Rayleigh length) while maintaining sufficient overlap with the interface to promote interaction with the FSG, without cracking it. The FSG-powder interface was taken as the reference plane. At the position defined as z0, the upper boundary of the calculated focal volume coincides approximately with this interface, so that the entire nominal focal depth (~1340 µm) extends into the powder bed. Positive displacements correspond to movement of the beam waist further into the powder, whereas negative displacements move it towards the FSG (see **Figure 3(E)**).

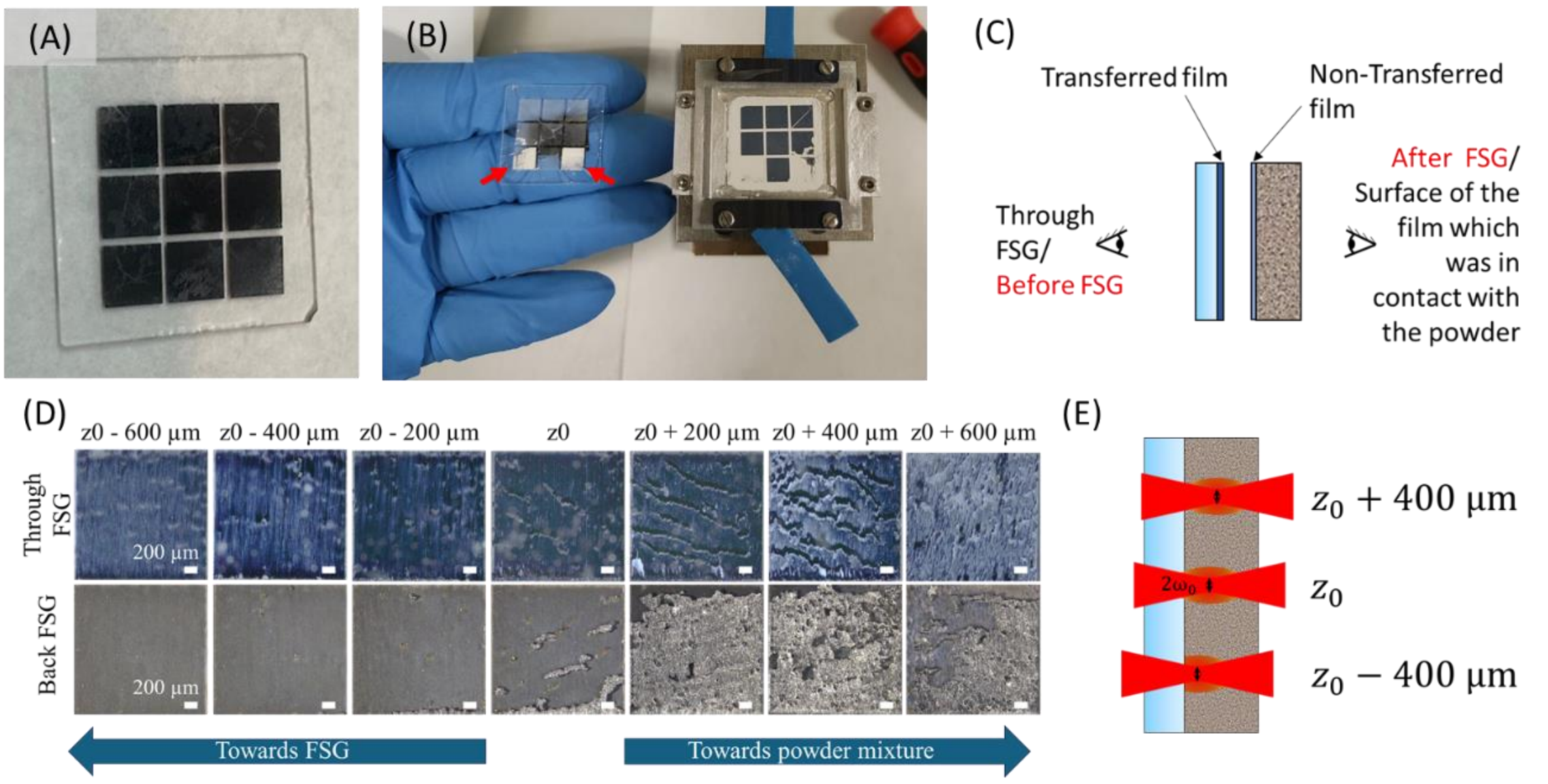

**Figure 3.** (A) Transference scenario (E = 4 µJ, v = 0.1 m/s, 6 × 6 $mm^2$ areas). (B) Non-transference scenario (E = 1 µJ, v = 0.1 m/s, 6 × 6 $mm^2$ areas); films that appear attached at the FSG (pointed with red arrows) can easily been removed from there by soft blowing. (C) Layout of the samples found after removing the FSG. (D) Effect of focal position on material transfer in the confined powder-bed configuration. The focal position was varied in 200 µm steps relative to z0. Upper row shows the optical micrographs of the layers through the FSG. Lower row shows the optical micrographs of the back side of FSG (in contact with the powder). (E) Layout of the beam Rayleigh range for three different positions in the z-search.

When the beam waist was displaced further into the powder bed, a larger amount of unprocessed powder remained associated with the transferred material (see **Figure 3(D)**, lower row). Conversely, displacement towards the FSG produced cleaner transferred layers (see **Figure 3(D)**, upper row). Within the investigated range, however, the transfer behaviour remained relatively stable, indicating a broad focal-position tolerance at this pulse energy. A focal position of z0 - 400 µm was therefore selected for the subsequent optimal energy search. Thus, most of the focal volume remained within the powder mixture while maintaining overlap with the FSG-powder interface (see **Figure 3(E)**). Since the laser beam propagates through the FSG and interacts with the powder bed, energy deposition is expected to occur predominantly within the powder region overlapped by the focal depth.

A second set of irradiation experiments was conducted to find the effect of the pulse energy in the resulting material, as illustrated in **Figure 4**. Eight 2 × 2 $mm^2$ areas were processed, increasing the pulse energy from 1 µJ to 8 µJ in steps of 1 µJ and with a scanning speed v = 0.10 m/s. In **Figure 4(A)**, these areas are framed inside of a yellow rectangle, showing them through the FSG just after the irradiation, whereas in **Figure 4(B)** the irradiations are shown on the FSG after being removed from the powders. Three different regimes can be identified by visual inspection, attending at the colour of the laser processed areas: E = 1 µJ (grey-bluish colour), 2 µJ ≤ E ≤ 5 µJ (blue colour), and E ≥ 5 µJ (white colour). In the first regime, the processed area shows a grey appearance, a bit darker than the powder colour, that turns blue in the second regime. This blue colouring is usually correlated to the reduction of Nb from $Nb^{5+}$ to $Nb^{4+}$, always associated to an increase of the oxide conductivity [23].

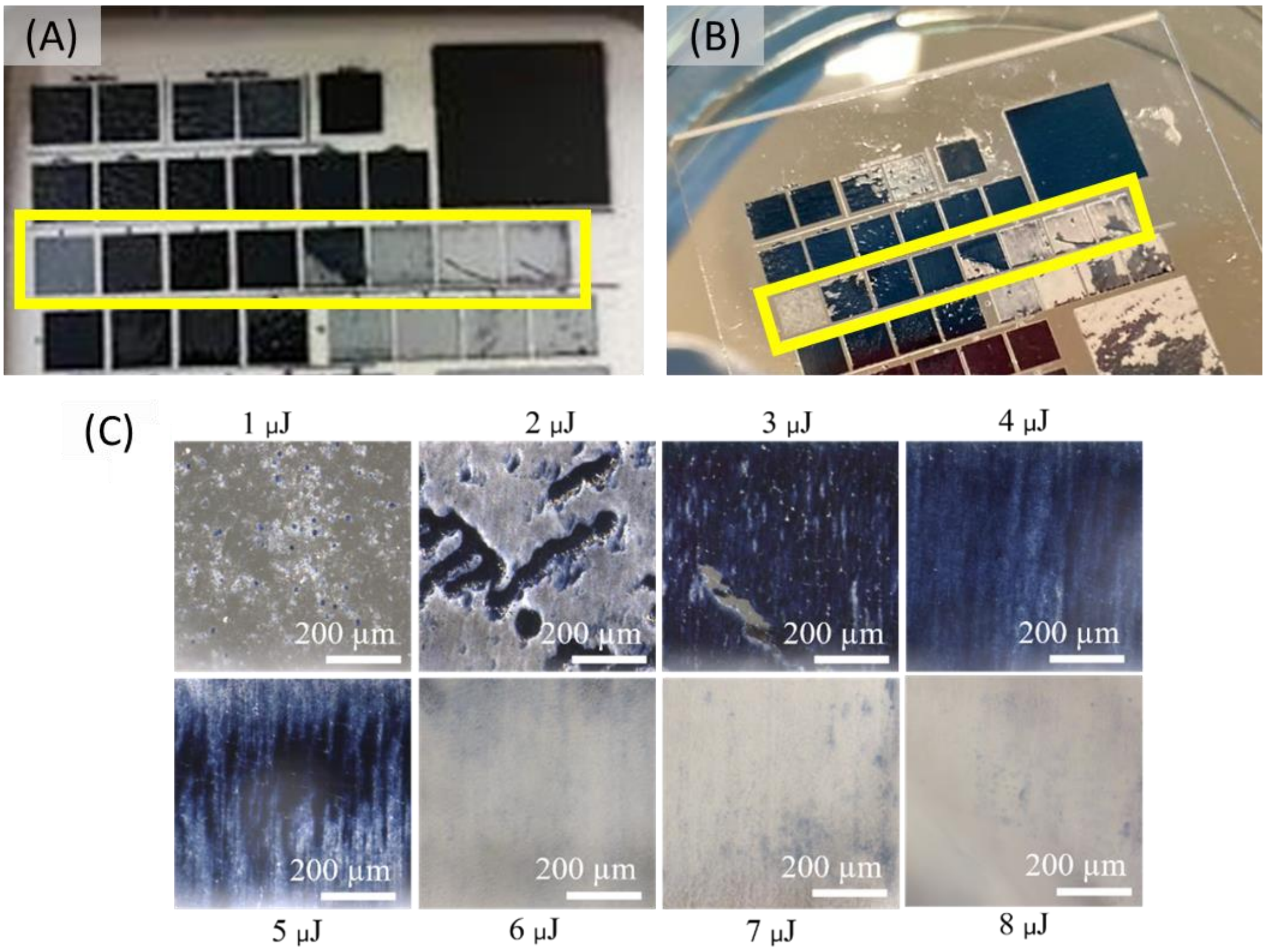

**Figure 4.** Appearance through the FSG of the laser-processed areas in E-study (A) before and (B) after removing the FSG from the irradiated powder. Only the arrays of 2 × 2 $mm^2$ areas highlighted by a yellow square belong to this study. (C) Optical micrographs (recorded trough the FSG) of the material transferred to the FSG after processing at E = 1 - 8 µJ. These areas are those highlighted with a yellow square in (A) and (B).

OM recorded at higher magnification for all the energies are presented in **Figure 4(C)**. These give some hints of how the material is transferred to the FSG. At 1 µJ, isolated micrometre-sized blue particles are observed on the FSG together with residual unmodified powder. Their rounded morphology is consistent with local melting and short-range material transport towards the FSG. These particles could be an initial stage of the formation of a continuous layer of material, because from energies higher than 2 µJ a continuous film is present on the FSG, as shown in **Figure 4(C)**. As the pulse energy increases, the amount of processed material retained on the FSG increases, evolving towards a continuous blue layer at 3 - 4 µJ. At 5 µJ, we see both blue and white regimes, as seen in **Figure 4(A-B)**, with different surface morphology (**Figure SM3**). A homogeneous white appearance is established from 6 µJ. As well as a thin film of processed material remains attached at the surface of the FSG for E = 2 - 8 µJ, there is also modified material that is left on the powder bed for E = 1 - 4 µJ (see **Figure 3(C)**). From the above results, three representative pulse energy conditions have been selected: 1 µJ (non-transference case), 4 µJ (transference scenario, blue colour regime) and 6 µJ (transference scenario, white colour regime).

### 3.2 Non-transferred material: low-energy regime.

Areas of both 2 × 2 mm$^2$ and 6 × 6 mm$^2$ were processed to study the properties and homogeneity of non-transferred layers produced under the laser processing conditions shown in **Table 1**. Larger areas provide enough material to perform XRD and electrochemical measurements. No differences have been observed upon changing the size of the irradiation area.

**Table 1.** Fs-laser processing conditions for low-energy regime, including the fluence at focus (F, corrected by the transparency of the lenses) and the number of effective pulses used ($N_{eff}$). F was calculated as $F = T(\%)\cdot(2\cdot E\cdot\pi^{-1}\cdot\omega_0^{-2})$, being T(%) the transmittance of the focusing lens at 1030 nm. $N_{eff}$ was calculated as $N_{eff} = \pi\cdot\omega_0^2\cdot f_{rep}\cdot v^{-1}\cdot d^{-1}$.

| E [µJ] | v [m·s$^{-1}$] | d [µm] | z position | F [mJ·cm$^{-2}$] | $N_{eff}$ [pulses] |
|---|---|---|---|---|---|
| 1 | 0.10 | 4 | z0 - 400 µm | 154 | 1344 |

**Figure 5** sums up the results observed by OM, SEM and EDX-SEM on the recovered non-transferred 1 µJ areas. **Figure 5(A)** shows the OM of the area, whereas **Figures 5(B-C)** show SEM images taken on the surface of the film. **Figure 5(B)** shows that the produced film presents some porosity, with microholes throughout the sample, and micrometric surface corrugations covered by micro- and nanocrystals. These structures vary in size and shape, including large (~1–2 µm) and small (<1 µm) spherical particles, as well as even smaller particles with varied geometrical shapes. They are smaller and differ in morphology from the original powder mixture (**Figure SM4**), and their abundance, particularly that of spherical particles, suggests a melting process involving the ejection of droplets. The estimated mean thickness of the 1 µJ layers is below 20 microns.

In **Figures 5(C-D)**, the corner of a 1 µJ processed squared-area is shown by BSE- and EDX-SEM. In **Figure 5(D)** the Ti (blue) and Nb (red) distribution maps measured by EDX appear superimposed (in Figure SM5 the individual EDX maps for each element are also available). Nb is homogeneously distributed in the film layer, contrary to what happens for the Ti case, which appears clustered in the left edge of the film, as well as along the layer. The average Ti:Nb atomic ratio is 1:5, consistent with the stoichiometry of $Ti_2Nb_{10}O_{29}$ instead of the target 1:2 atomic ratio of $TiNb_2O_7$. Thus, the sintered layer experienced a Nb-enrichment upon the fs-laser processing (the pristine powder presented a 1:2 atomic ratio).

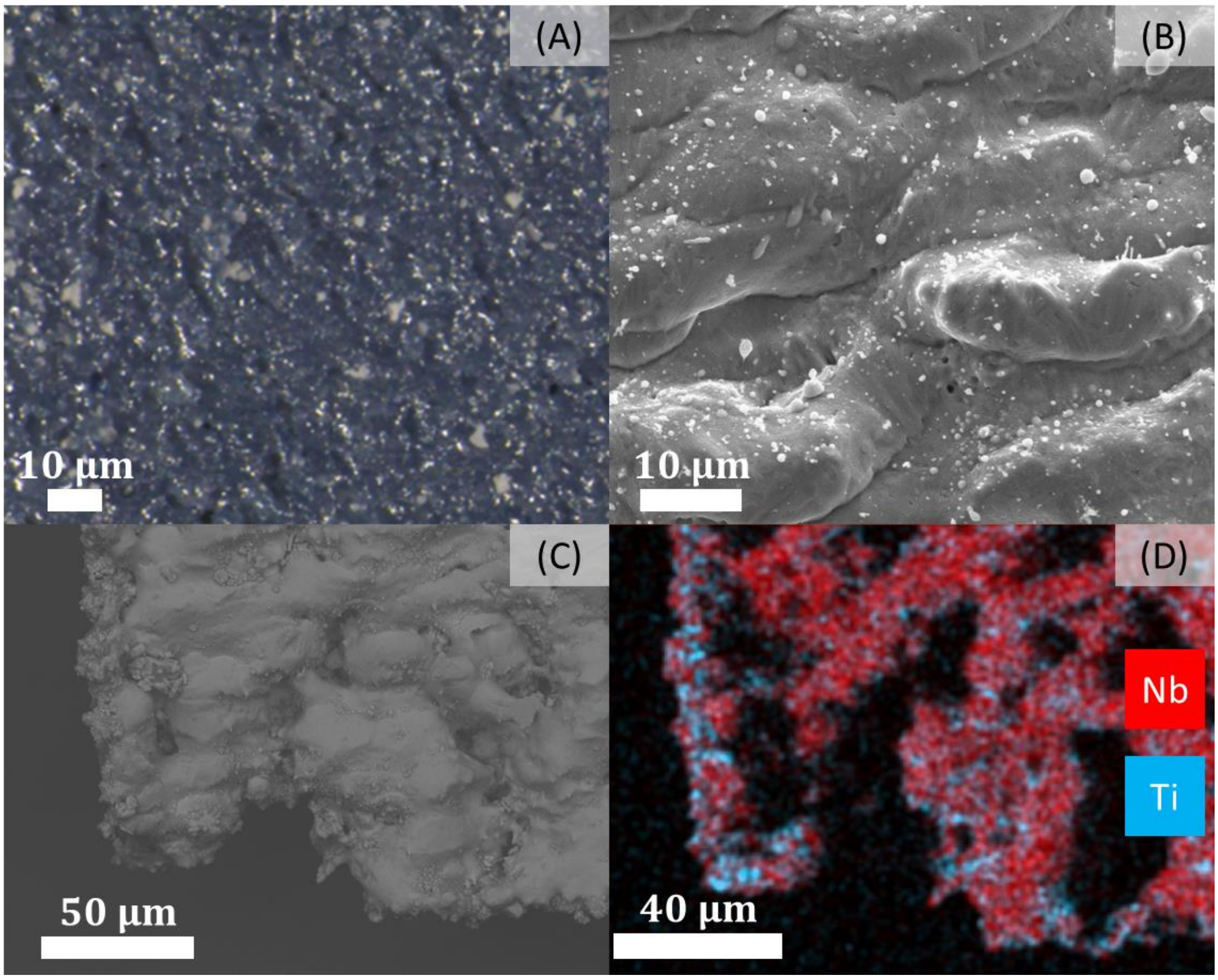

**Figure 5.** (A) OM picture of low-energy regime area (1 µJ, 0.10 m/s). (B) SE-SEM top-view picture of a film recovered from the powder, focused on the face of the layer that was in contact with the FSG before removing it. (C) BSE- SEM picture of one corner of a non-transferred layer recovered from the powder. (D) EDX-SEM maps on image (C), based on the Lα-lines intensity of Nb (red) and Kα-lines intensity of Ti (blue).

To confirm the TNO phase produced, gonio-mode XRD and top-view micro-RS studies have been conducted. The acquired measurements are shown in **Figure 6**. Both XRD and RS studies indicate the presence of a highly crystalline 1:5 TNO (i.e. $Ti_2Nb_{10}O_{29}$). The reflections present in the reference monoclinic phase of $Ti_2Nb_{10}O_{29}$ diffractogram show a good agreement with the collected pattern, as seen in **Figure 6(A)**. The most intense reflections match those of the reference in position and relative intensity in PDF: #01-072-0159. Due to the sharpness and good definition of the XRD reflections, a relatively large grain size of the TNO is expected for the laser-sintered material, as it has been reported for other solid-state produced TNOs [7].

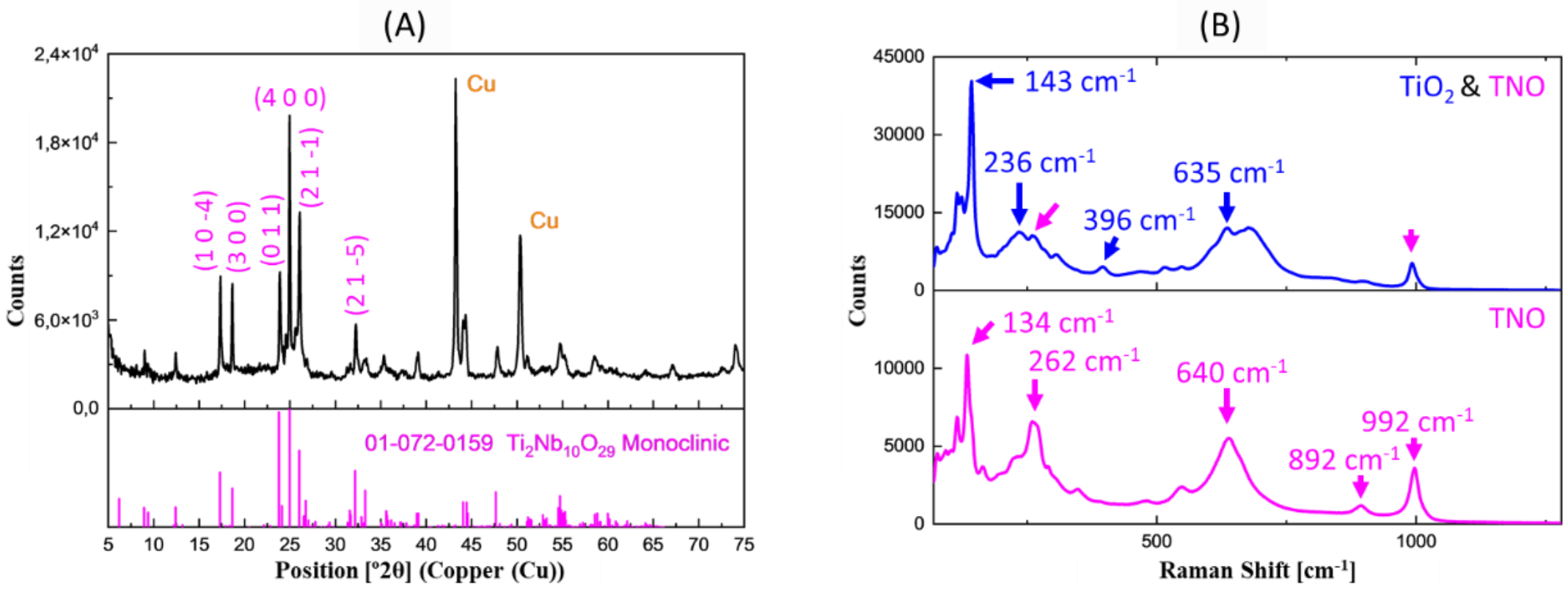

**Figure 6.** (A) XRD measurement of a 1 µJ area recovered from the powder (black) and XRD pattern of a monoclinic $Ti_2Nb_{10}O_{29}$ (pink) taken from reference PDF: #01-072-0159. The most intense reflections are labelled according to the reference. Two intense reflections related to the Cu substrate used are marked in the measurement according to reference PDF: #00-004-0836. (B) Micro-RS measurements performed in two different places of the same 1 µJ

recovered film. The coloured arrows indicate the presence of relevant Raman modes related to TNO (pink) and $TiO_2$ (blue). The pink arrows in the upper spectrum refer to, respectively, 260 and 990 $cm^{-1}$ (TNO modes).

However, as the XRD measurement averages over the spotsize of the probe (2.5 × 4.5 $mm^2$), micro-RS measurements have also been performed at several positions on the film. Two different Raman spectra have been identified, shown in **Figure 6(B)**. First, the most representative one is shown at the bottom of **Figure 6(B)**, the pink spectrum, which confirms the formation of a crystalline TNO phase [7,24], in agreement with the XRD measurements. Both the 1:2 and 1:5 TNO stoichiometries show a similar Raman spectrum [24], however, XRD is clear at this respect: $Ti_2Nb_{10}O_{29}$ is the laser-sintered stoichiometry. On the other hand, the blue spectrum at the top of **Figure 6(B)** shows the characteristic Raman modes of TNOs (pink arrows) together with $TiO_2$-related vibrations from both anatase [25,26] and rutile [27] (blue arrows), although the precursor $TiO_2$ powder was solely anatase. The intense anatase modes probably arise from $TiO_2$ powder transferred to the sintered layer during recovery; otherwise, temperatures high enough to trigger the irreversible anatase-to-rutile transition would suppress anatase peaks (see the rutile Raman mode at 236 $cm^{-1}$ [27]). These results point to the formation of a $Ti_2Nb_{10}O_{29}$:$TiO_2$ composite upon fs-laser treatment of the precursors at 1 μJ.

Finally, when moving to the non-transferred material produced at pulse energies ranged 2 - 4 μJ (which is present on the powder bed regardless of whether there is or not transference to the FSG), we find a more complex composition than for E = 1 μJ. XRD measurements performed on new non-transferred material show that, although a change in coloration is observed after irradiation, the sintered material is composed by a mixture of many oxides: both H-$Nb_2O_5$ and T-$Nb_2O_5$, anatase $TiO_2$ and a portion of monoclinic $Ti_2Nb_{10}O_{29}$ (see **Figure SM6**). We have decided to focus our application study on the 1 μJ pulse energy only (see **Section 3.5**), as this is the only case where the material is predominantly non-transferred, and it is also the only system with relatively low compositional complexity.

### 3.3 Transferred layers

The morphology evolution of the material transferred in 2 × 2 $mm^2$ areas to the FSG substrate is shown in the SEM images in **Figure 7.** As described above, for the 1 μJ condition, only droplets of melted material are observed on the FSG (see **Figure 7(A)**). For energies above 3 μJ, continuous layers are formed on the FSG, as shown in **Figures 7(B-C)**. The conditions selected as representative of continuous-layer formation are therefore 4 μJ and 6 μJ, as described in the following subsections.

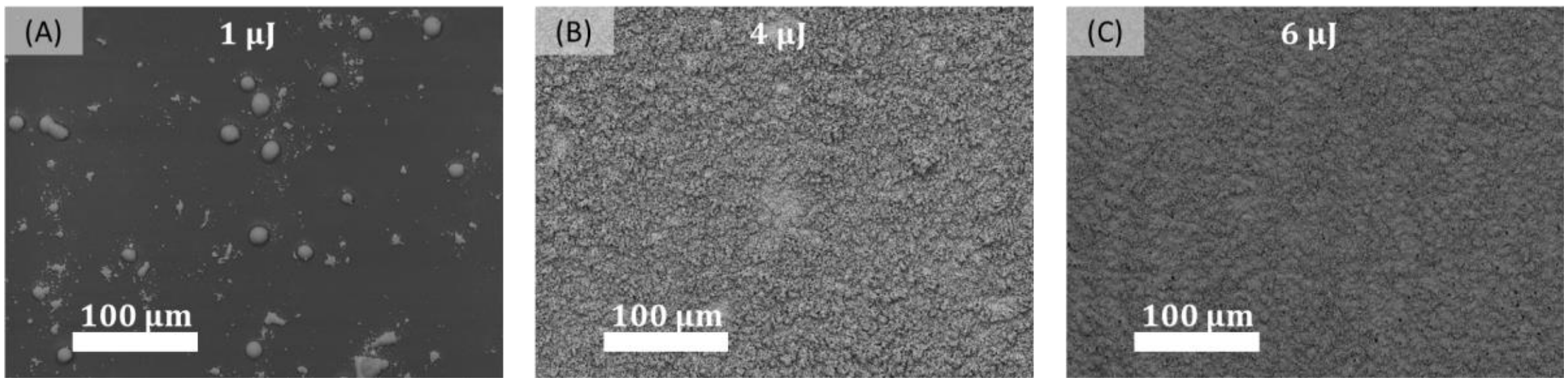

**Figure 7.** SE-SEM images of the (A) 1, (B) 4 and (C) 6 μJ areas transferred to the FSG.

#### 3.3.1 Intermediate-energy regime: 4 μJ.

As previously, both 2 × 2 $mm^2$ and 6 × 6 $mm^2$ areas were subjected to detailed characterization. The processed conditions employed are listed in **Table 2**. No appreciable differences in film appearance or morphology were observed between the two area sizes.

**Table 2.** Fs-laser processing conditions for intermediate-energy regime, as presented for Table 1.

| E [μJ] | v [m·$s^{-1}$] | d [μm] | z position | F [mJ·$cm^{-2}$] | $N_{eff}$ [pulses] |
|---|---|---|---|---|---|
| 4 | 0.10 | 4 | z0 - 400 μm | 619 | 1344 |

**Figure 8** summarizes the results obtained by OM and SEM. **Figure 8(A)** shows by OM the colour and appearance of a transferred 4 µJ film when it is observed through the FSG (upper half of the image) and when it is directly observed (bottom half). **Figure 8(B)** shows by BSE-SEM the surface morphology of this naked face of the sintered film. We can distinguish larger spherical features than those observed in **Figure 5** and that have been, apparently, melted, deposited and re-solidified onto the film.

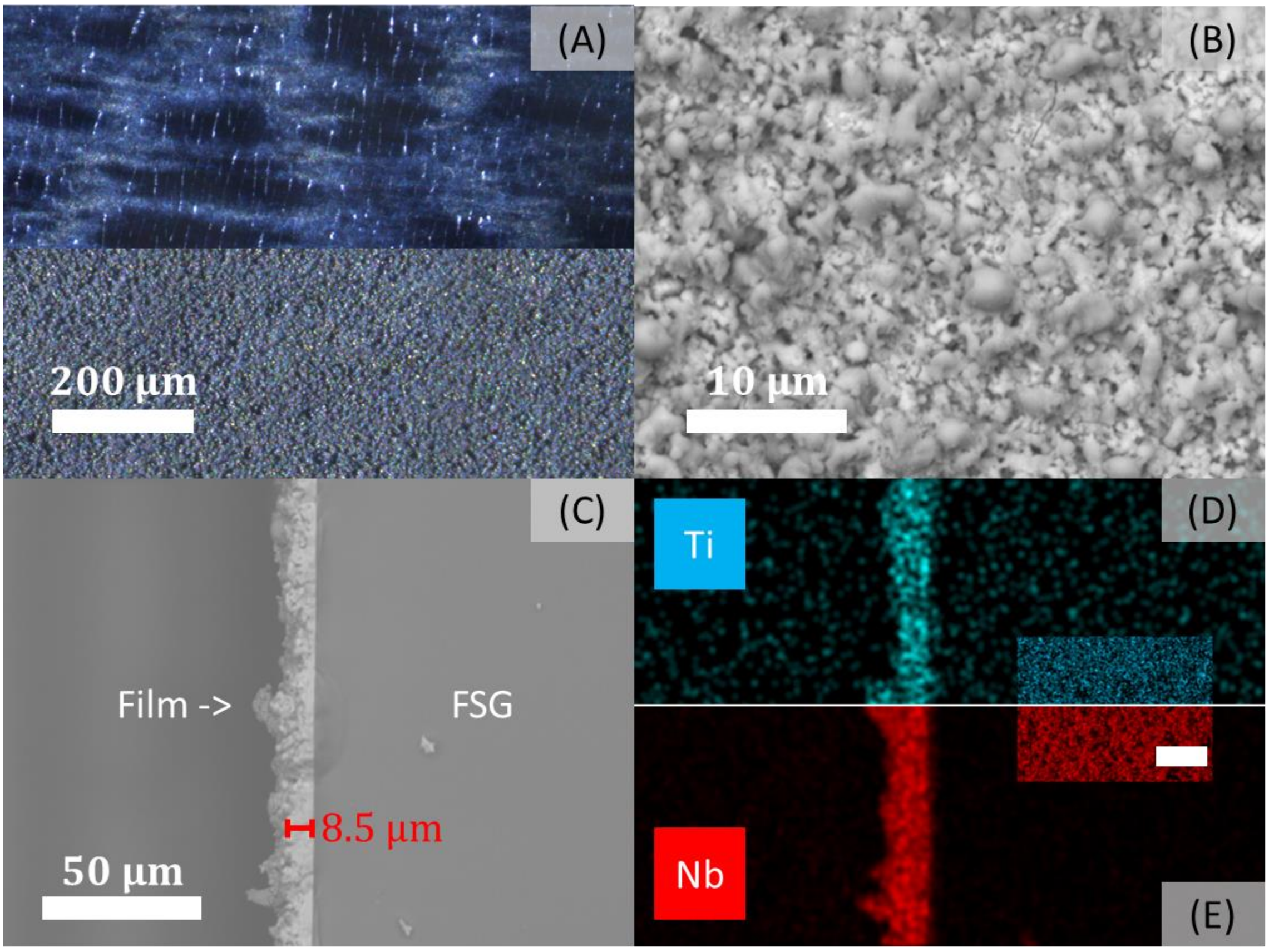

**Figure 8.** (A) Top-view OM pictures of a transferred intermediate-energy regime film (4 µJ, 0.10 m/s). The picture shows the appearance and colour of the film when directly observing its surface (top half) and when observing through the FSG (bottom half). (B) SEM top-view picture of the area, focused on the face of the layer that was in contact with the powder before removing the FSG. (C) X-SEM of the same area, showing the thickness of the layer (red) (D) Nb- and (E) Ti-distribution maps measured by EDX-SEM on image (C). The insets show the equivalent map for a top-view, with a white scalebar of 50 µm. The EDX results are based on the Lα-lines of Nb and Kα-lines of Ti.

The films were cleaved to measure their thickness and composition in depth. These results appear in **Figure 8(C-E)**. From X-SEM measurements, the thickness was estimated to be 8.5 ± 1.0 µm, considering the film roughness to estimate the error. There is no sign of the presence of a gradient in depth for the distribution of Nb or Ti, which appear uniformly distributed in the whole thickness of the film. In the insets of both EDX maps it is shown that the superficial distribution is also homogeneous, contrary to what happened in the 1 µJ non-transferred film. For the 4 µJ condition, the atomic Ti:Nb ratio, measured by EDX is 1:4, which is also not consistent with the desired $TiNb_2O_7$ stoichiometry.

Gonio-mode XRD and X-RS studies have been performed to analyse the phases present in the films. XRD is shown in **Figure 9(A)**, where the measured pattern (blue colour) presents a good matching with the (0 0 1) and (1 0 0) reflections of pseudo-hexagonal TT-$Nb_2O_5$ (PDF: #00-028-0317), as well as with the (-1 1 0) and (-3 0 3) reflections of monoclinic phase $TiNb_2O_7$ (PDF: #00-039-1407). Apart from these reflections, other less intense maxima can be identified as crystallographic planes of the same oxides. The presence of TT-$Nb_2O_5$ together with $TiNb_2O_7$ justifies the Nb enrichment seen by EDX.

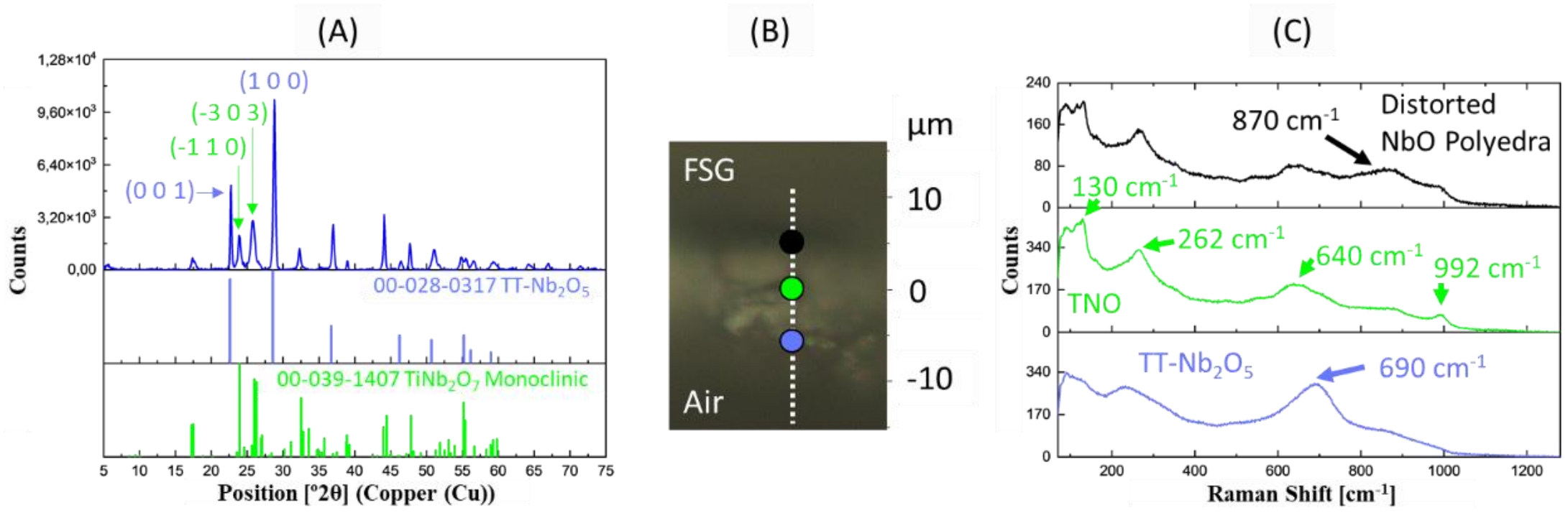

**Figure 9.** (A) XRD measurement of a 4 µJ area (blue). Underneath, XRD patterns of TT-$Nb_2O_5$ (violet) and monoclinic $TiNb_2O_7$ (green) taken from references PDF: #00-028-0317 and #00-039-1407, respectively. The more intense reflections of the measured area are labelled according to the colour of the reference diffractograms. (B) X-OM picture of the same area. The three coloured dots show the places of the film in which the (C) micro-RS measurements were performed: close to the film/FSG interface (black), the middle part of the film (green) and close to the air/film interface (blue). The coloured arrows indicate the presence of relevant Raman modes related to TNO (green), TT-$Nb_2O_5$ (violet) and [$NbO_6$] polyhedra (black).

**Figures 9(B) and 9(C)** show the X-RS of the 4 µJ film. Raman analysis has been performed on three different regions marked by the three coloured dots present in **Figure 9(B)**. In **Figure 9(C)**, the measured spectra are plotted preserving the colours used in the OM image. Close to the FSG, the material shows a more distorted Nb-O environment, as indicated in the broad Raman band located around 870 $cm^{-1}$ in the black spectrum, generally associated with highly distorted [$NbO_6$] polyhedra [28]. However, in the middle of the sintered film, the material becomes crystalline and shows a spectrum compatible with the formation of a monoclinic $TiNb_2O_7$, as proved by XRD. The absence of $TiO_2$ related peaks (either anatase or rutile) agrees with the good incorporation of Ti into the $Nb_2O_5$ structure. Finally, at the interface between the TNO and the air, the Raman spectrum is consistent with that of a TT-$Nb_2O_5$ structure [29]. This means that a layered structure is formed, with the targeted $TiNb_2O_7$ film which is sandwiched between the FSG and a Nb-enriched layer formed behind the TNO.

### 3.3.2 High-energy regime: 6 µJ.

At a pulse energy of 6 µJ, the irradiation conditions are summarised in **Table 3**. The transferred layers were first examined in 2 × 2 $mm^2$ areas before extending the processing to 6 × 6 $mm^2$ areas.

**Table 3.** Fs-laser processing conditions for high-energy regime, as presented for Table 1.

| E [µJ] | v [$m \cdot s^{-1}$] | d [µm] | z position | F [$mJ \cdot cm^{-2}$] | $N_{eff}$ [pulses] |
|---|---|---|---|---|---|
| 6 | 0.10 | 4 | z0 - 400 µm | 926 | 1344 |

The resulting 2 × 2 $mm^2$ areas showed continuous transferred layers with a predominantly white appearance, as described in **Section 3.1** and shown in **Figure 4(C)**. The X-SEM image indicates a transferred-layer thickness of 20 ± 1.0 µm (**Figure 10(A)**), greater than that measured for the 4 µJ layers. X-RS measurements (**Figure 10(B-C)**) indicated the formation of $TiNb_2O_7$ together with an Nb-enriched outer region, similarly to the layers produced at 4 µJ.

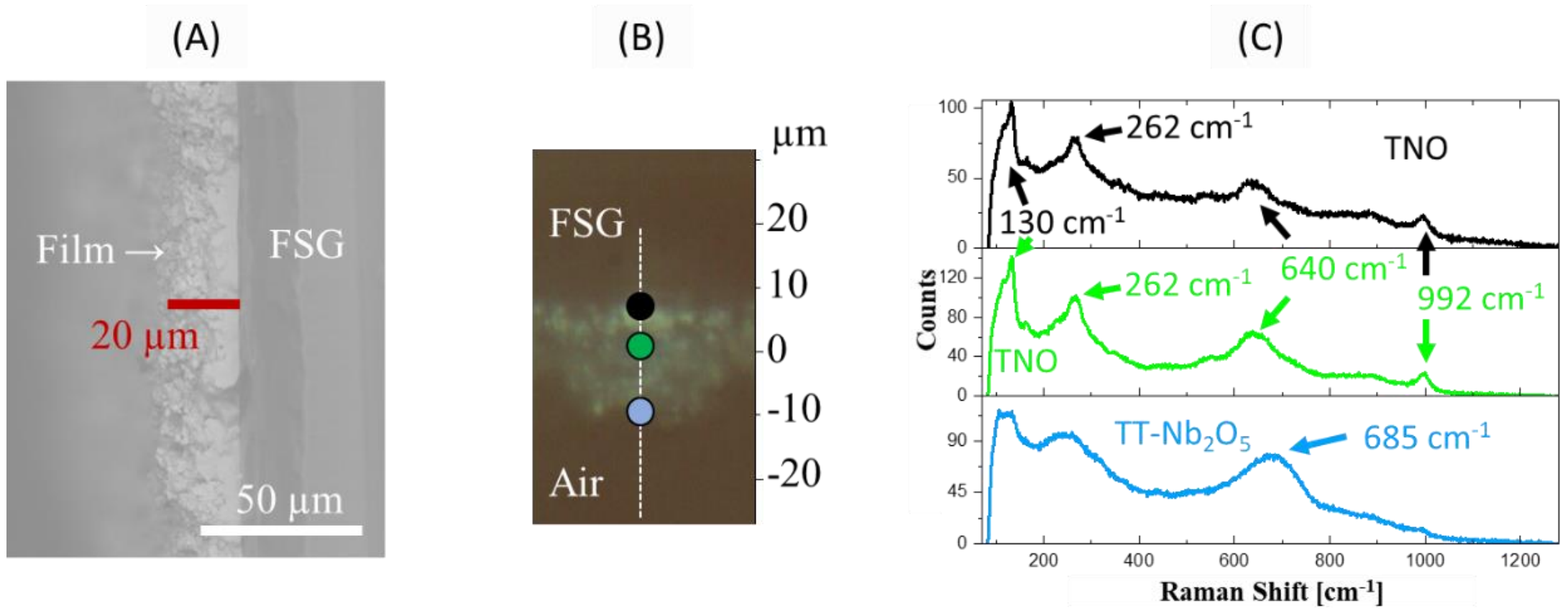

**Figure 10.** Characterization of a 2 × 2 $mm^2$ layer (E = 6 μJ, v = 0.10 m/s). (A) X-SEM image of the layer transferred onto the FSG, showing its thickness (red). (B) X-OM of the same layer. The three coloured dots indicate the positions at which the micro-Raman spectra in (C) were recorded: close to the film/FSG interface (black), in the middle of the film (green), and close to the powder-facing surface (blue). The Raman features are comparable to those described in **Figure 9**.

As for the intermediate-energy regime, 6 × 6 $mm^2$ areas were subsequently processed to obtain a sufficiently large area for XRD and four-point-probe electrical measurements. However, maintaining the conditions used for the 2 × 2 $mm^2$ areas resulted in a loss of layer quality, with less continuous films and the coexistence of blue and white regions. Thus, the conditions that produced satisfactory small-area layers did not provide comparable uniformity when the processed area was increased.

This behaviour motivated additional optimization of the focal position and scanning speed. A focal-position study was performed at the transition pulse energy of 5 μJ discussed in **Section 3.1** (**Figure SM7**). Moving the beam waist towards the FSG promoted the white phase appearance, although cracking of the substrate also became apparent. A focal position of z0 - 600 μm was therefore selected for the subsequent 6 μJ irradiations. The scanning speed was also reduced to 0.05 m/s, increasing the number of pulses delivered per unit length and the nominal energy input per scanned area. The processing conditions used to produce the characterized 6 × 6 $mm^2$ layers are listed in **Table 4**.

**Table 4.** Adjusted fs-laser processing conditions for high-energy regime, as presented for Table 1.

| E [μJ] | v [$m \cdot s^{-1}$] | d [μm] | z position | F [$mJ \cdot cm^{-2}$] | $N_{eff}$ [pulses] |
|---|---|---|---|---|---|
| 6 | 0.05 | 4 | z0 - 600 μm | 926 | 2688 |

The sensitivity to the processed area may be related to changes in heat accumulation: for longer scanning lines, the beam takes more time to reach neighbouring regions again. The need to adjust both the focal position and scanning speed is consistent with a dependence on local energy accumulation near the FSG-powder interface. These adjustments allowed larger transferred layers to be obtained for structural and electrical characterization, although their morphology and appearance remained nonuniform (as shown in **Figure SM8(A)**). We also confirmed that both Nb and Ti were transferred to the FSG using EDX-SEM measurements (**Figure SM8(B-C)**).

XRD measurements performed on these areas are shown in **Figure 11(A)**, indicating a good matching with the main reflections of monoclinic phase $TiNb_2O_7$ (PDF: #00-039-1407). **Figures 11(B) and 11(C)** show the X-RS study. The black spectrum presents the usual fused silica bands [30]. At the FSG-layer interface, the red spectrum shows a weak peak at approximately 994 $cm^{-1}$, associated with Nb-O vibrations in the crystallographic shear structures of TNOs [7,25]. The spectrum recorded in the inner part of the layer (green) shows good agreement with the expected $TiNb_2O_7$ spectrum [7,24], consistent with the XRD identification. Finally, the powder-facing side of the transferred layer (turquoise) also shows a TNO-like spectrum, although with lower intensity and greater noise due to the vicinity of the interface with air. Micro-RS of 2 × 2 mm$^2$ areas obtained with the adjusted conditions in **Table 4** show similar behaviour. However, as expected, the homogeneity of the film (thickness and colour) is preserved in the smaller areas.

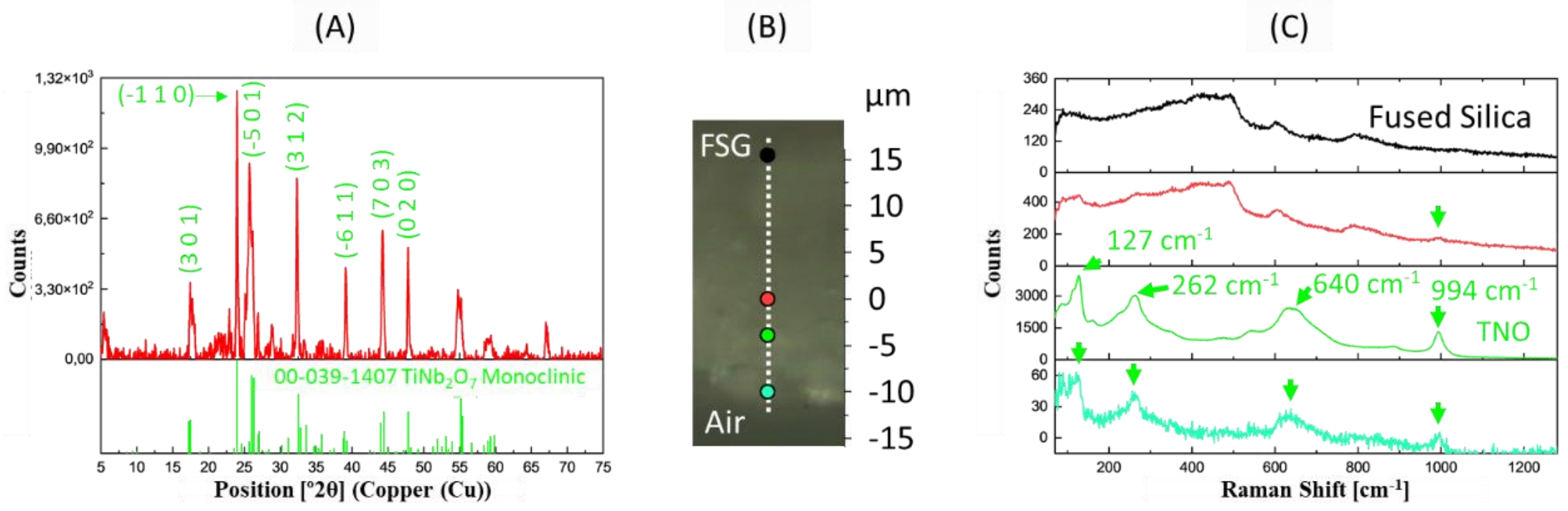

**Figure 11.** (A) XRD measurement of a 6 µJ, 0.05 m/s area. Underneath, XRD pattern $TiNb_2O_7$ (green) taken from reference PDF: #00-039-1407. The most intense reflections are labelled according to the reference. (B) Cross OM picture of the same 6 µJ area. The four coloured dots show the places of the film in which the (C) micro-RS measurements were performed.

Overall, both the intermediate- and high-energy processing conditions produced layers containing $TiNb_2O_7$. The adjusted high-energy conditions yielded a layer without detectable secondary crystalline phases by XRD, but with substantially poorer thickness uniformity in larger areas than the 4 µJ layers conditions. Nevertheless, the optimization was successful in enabling the formation of layers containing the target $TiNb_2O_7$ phase.

### 3.4 Proposed layer-formation mechanism.

After producing and characterizing the fs-laser sintered films for the three energy regimes described, different aspects of the TNO formation mechanism can be discussed. In **Table 5**, a summary of the films' characteristics achieved for each energy regime is presented. There is an evident trend in terms of the achieved stoichiometry.

For the lowest pulse energy, as it was discussed in **Section 3.2**, the formation of $Ti_2Nb_{10}O_{29}$ together with the presence of residual $TiO_2$ indicates that the precursor mixture did not undergo a complete transformation into the targeted $TiNb_2O_7$ phase. However, this agrees with previous studies, in which $Ti_2Nb_{10}O_{29}$ is reported to appear in the initial stages of the formation of $TiNb_2O_7$ from the oxides [31]. The structural characterization by XRD and RS, in combination with the appearance of the droplets covering the low-energy film (**Figure 5(C)**) and the FSG (**Figure 7(A)**), reinforces the idea of a melting and re-solidification process. $Ti_2Nb_{10}O_{29}$ can be formed at 1478 ºC through the congruent reaction of Liquid→$Ti_2Nb_{10}O_{29}$ [3], and the melting points of our powder precursors are 1490ºC and 1825ºC ($Nb_2O_5$ and $TiO_2$, respectively). Accordingly, we deduce that film formation from the precursor powders upon accumulation of 1 µJ pulses may involve transient high temperatures, with at least partial melting and subsequent re-solidification. However, these temperatures are not sustained long enough to promote complete Ti diffusion into the niobium oxide lattice.

**Table 5.** Types of fs-laser sintered layers under the three group of laser conditions used. The full description of each regime is shown in **Tables 1, 2 and 4**.

| Laser-sintering conditions regime | Physical substrate | Film colour | Identified phases |
|---|---|---|---|
| Low-energy<br>(E = 1 µJ, v = 0.10 m/s) | Free-standing<br>(Non-transferred) | Grey/Blue | $TiO_2$ (anatase)<br>$Ti_2Nb_{10}O_{29}$:$TiO_2$ (rutile) |
| Intermediate-energy<br>(E = 4 µJ, v = 0.10 m/s) | FSG<br>(Transferred) | Blue | $TiNb_2O_7$<br>TT-$Nb_2O_5$ |
| Adjusted high-energy (E = 6 μJ, v = 0.05 m/s) | FSG (Transferred) | White | $TiNb_2O_7$ |

For the intermediate-energy conditions, transference of continuous layers of the targeted $TiNb_2O_7$ to FSG is produced, although TT-$Nb_2O_5$ remains detectable. Again, the observed droplets and transferred-layer morphology of this scenario support a mechanism involving partial melting, short-range transport towards the FSG and subsequent re-solidification. In this case, the accumulation of 4 μJ pulses appears to generate temperatures sufficiently high and sustained to promote complete Ti diffusion into the niobium oxide lattice. Broader XRD reflections of the 4 μJ films compared with the 1 μJ material may indicate a smaller coherent crystallite size and/or greater microstructural strain, coherent with the different solidification rates described. Then, because of the presence of the FSG, the liquid quenches faster and the congruent reaction of Liquid→$TiNb_2O_7$ occurs [3]. It must be considered that the transition temperatures reported by C. Liu et al. [3] for the mentioned congruent reactions (i.e. 1491 °C) were calculated for a different precursors' molar ratio than the one we are using in our experiments. Therefore, the temperatures reported in the equilibrium phase diagram shown in [3] should only be considered as reference values for the present laser-processing conditions.

Furthermore, X-RS identifies a $Nb_2O_5$-rich region on the powder-facing side of the 4 μJ layers, suggesting spatial variation in the extent of TNO formation (see **Figure 9**). The remaining material, which is not transferred but stays in the powder bed, contains a mixture of $Nb_2O_5$, $TiO_2$ and $Ti_2Nb_{10}O_{29}$ (see **Figure SM6**). The persistence of TT-$Nb_2O_5$ on the powder-facing side of the transferred layer, together with the composition of the non-transferred material remaining in the powder bed, suggests a gradient in reaction extent through the processed volume. Preferential energy deposition near the FSG-powder interface, combined with differences in heat transfer and material redistribution, may contribute to less complete TNO formation in regions further from the interface.

Finally, under the adjusted high-energy conditions listed in **Table 4**, only the targeted $TiNb_2O_7$ stoichiometry is detected, without additional secondary phases. However, the process appears to be highly dependent on thermal accumulation, making the production of larger homogeneous areas more challenging. Regarding the formation mechanism, the 6 μJ laser pulses trigger an intense heating of the mixture of powders, producing the melting and ejection process previously proposed, but at a higher temperature. When the molten material reaches the FSG at high temperature, it undergoes a rapid thermalization and quenches, triggering the formation of an initial TNO thin film with poor crystallinity. This is evidenced by the simultaneous presence of FSG bands and the slight 994 $cm^{-1}$ mode in the Raman spectra of **Figure 11**. A similar behaviour was observed in the intermediate-energy layers, particularly in the region of the film closest to the FSG (see **Figure 9(C)**). After the formation of the initial layer, the hot liquid ejected from the powder bed reaches this film instead of the FSG, crystallizing better on the TNO seed and forming the desired material on top.

In summary, the proposed mechanism is similar to a Laser Induced Backward Transfer mechanism (LIBT) [32], in which the laser triggers the transference of material upon melting of a solid target and ejection of the liquid material, that is deposited on a transparent substrate placed between the laser source and the target material. In contrast to usual LIBT technique, we are using a mixture of powders as a precursor, instead of the solid material that we desire to transfer to a transparent substrate. A.A. Voskanyan et al. [15] demonstrated that the formation of both $TiNb_2O_7$ and $Ti_2Nb_{10}O_{29}$ stoichiometries by solid-state reaction reported to become thermodynamically favourable at elevated temperatures (around 1300 ºC and 1100 ºC, respectively) because they have positive enthalpies of formation from the binary oxides used in this work ($TiO_2$ and $Nb_2O_5$). This agrees with the proposed formation mechanism, which relies on the powerful heating induced by the high repetition rate scanning of the mixture of powders, that allows the melting of the precursors and the later re-solidification of a TNO.

Furthermore, unlike conventional methods used to produce TNOs, which typically require prolonged exposure to high temperatures (as discussed in **Section 1**), the technique employed here is relatively energy-efficient. For instance, the most typical method used to produce TNOs is based on muffle furnaces which can heat a pellet prepared with the precursors during long periods (in the order of tens of hours) at temperatures above 1000 ºC. While this kind of furnaces present a generally high-power consumption rate, the fs-laser we are using has low consumption (< 1 kWh, if considering continuous working). Together with the speed of the irradiations and the scalability of the process moving to larger focal length lenses, the proposed technique seems a cost-effective alternative to the conventional solid-state procedures.

### 3.5 Device integration.

One of the most common applications for TNOs is their use as anodes for energy storage, in particular for Li-ion batteries. However, a noticeable drawback of using TNOs as electrodes for LIBs relies on their poor electronic and ionic conductivities [33,34]. In this sense, a good way to test the potential capability of a TNO layer to be integrated into this kind of devices would be measuring their conductivity. To do this, we must have in mind that the different processing regimes described above do not only determine the structure and properties of the TNO, but also the way in which the processed material can be integrated into functional devices. While the intermediate- and adjusted high-energy regimes produce continuous conductive layers that are directly supported on the FSG, the low-energy regime provides recoverable material which is not supported by any rigid substrate that could hold the layer while measuring without depositing electrodes. Due to this reason, transferred samples have been measured by means of electrical 4-point probing (**Section 3.5.1**), whereas the non-transferred material has been tested in an electrochemical cell (discussed in **Section 3.5.2**).

#### 3.5.1 Electrical measurements of transferred layers.

For the intermediate-energy conditions (4 μJ, 0.10 m/s), $R_\square$ of twenty different 6 × 6 $mm^2$ areas have been measured as described in **Section 2.4**. A mean thickness of 8.5 microns (extracted from the X-SEM performed, see **Section 3.3**) was used to calculate the resistivity of the samples. The results are presented in **Table 6**. Even though the errors associated to the measured magnitudes are relatively low, there is still some variability when changing from one area to another (see the Supplementary Material File, **Table SM1**), probably associated to the roughness shown by the films.

**Table 6.** Mean $R_\square$ and ρ obtained from the measurements of the intermediate-energy regime samples (4 μJ, 0.10 m/s). The error accompanying the mean values have been propagated as $20^{-1} \cdot \sum_1^n \sqrt{\Delta(x_n)^2}$, with $n$ referring to each set of 100 measurements and where $\Delta(x_n)$ is the experimental error measured for the $n$-set.

| Mean $R_\square$ [$\Omega \cdot \square^{-1}$] |
|---|
| (2242.2 ± 1.4) ·$10^2$ |
| **Mean ρ [Ω·cm]** |
| 190.58 ± 0.12 |

The achieved value of ρ = (190.58 ± 0. 12) Ω·cm, or in conductivity units σ ~ $5.25 \cdot 10^{-3}$ $S \cdot cm^{-1}$, is not only comparable, but enhances the typical resistivity values found in the literature for other Nb-based oxides and TNOs before lithiation. For example, J.M. Sieffert *et al.* reported conductivity values of $4.03 \cdot 10^{-9}$ $S \cdot cm^{-1}$ and $1.78 \cdot 10^{-9}$ $S \cdot cm^{-1}$ for, respectively, undoped and Nd-doped sol-gel TNOs [35]. R. Inada *et al.* reported a huge enhancement of the electrical conductivity of solid-state prepared TNOs when annealing them in vacuum instead of in air, showing that σ increased from $0.95 \cdot 10^{-11}$ $S \cdot cm^{-1}$ to $1.1 \cdot 10^{-3}$ $S \cdot cm^{-1}$ [36]. The result achieved in the work by Inada is in the same order of magnitude than the one achieved here (and still, the resistivity of our laser-sintered TNO is around 4 times lower), but there is an important difference in terms of energy and time cost that must be considered: the solid-state route requires a 24 hours annealing process at 1050 ºC, while the fs-laser sintering approach takes less than 4.5 minutes to produce a 1 $cm^2$ area of TNO. Moreover, the blue colour (see **Figure 8**) and good electrical conductivity of these films are consistent with the presence of reduced Nb species and/or oxygen-deficient states reported for TNOs [23,37], which would support the previously proposed formation mechanism. In summary, our intermediate-energy regime films show great potential for integration into a LIB.

On the other hand, the adjusted high-energy white areas (6 μJ, 0.05 m/s) were also measured by 4-point-probing, but the resistivity of the samples was high enough to overcome the limited measuring range of the equipment ($R_{\square}$ ~ 10 MΩ·$\square^{-1}$). However, differences in microstructure, porosity, grain boundaries and layer continuity may also contribute, and the present measurements do not allow these effects to be separated. Due to this, we have used the smaller 2 × 2 $mm^2$ areas to measure the resistivity just using 2-point-probing. Again, the measurement was overcoming the range of the equipment. This can serve as a qualitative observation that the material produced with 6 μJ is more resistant that the material obtained with 4 μJ. These results agree with previous studies, were TNO with oxygen-deficiency and blue-black coloration shows improved electrical conductivity [18,37].

For both energy regimes, a way to separate the TNO from the FSG to which it is attached must be found. Alternatively, other applications might beneficiate from the presence of a transparent substrate (photocatalysis, for instance), and the use of a conductive substrate that may allow the direct integration of the laser-sintered films into a device is also worth to study.

#### 3.5.2 Electrochemical performance of non-transferred 1 μJ films.

As the low-energy films are completely recoverable but too fragile to be measured with physic probes, we have used them to prepare a slurry (see **Section 2**) and measured their electrochemical performance to give some insights of their potential for LIBs.

Cyclic voltammetry (CV) was used to examine the redox behaviour (**Figure 12(A)**). At a scan rate of 0.1 mV/s, the first cycle differs noticeably from the subsequent scans. The first cathodic scan shows a broad reduction feature centred at approximately 1.18 V, which shifts towards 1.52 V from the second cycle, being consistent to $Nb^{5+}/Nb^{4+}$ reduction. A similar first-cycle evolution has been reported for TNOs and has been associated with irreversible changes occurring during the initial lithiation [7,38]. The anodic response is dominated by a broad peak around 1.86 V. Since individual CV features in defective $Ti_2Nb_{10}O_{29}$ have been reported to involve simultaneous Ti- and Nb-centred redox processes [38], this broad response is attributed to overlapping $Ti^{3+}/Ti^{4+}$ and $Nb^{4+}/Nb^{5+}$ oxidation contributions. The simultaneous presence of Ti- and Nb-related redox features is consistent with the formation of the TNO phase identified by XRD and RS. From the second cycle onwards, the voltammograms show a high degree of overlap, with only minor changes in peak position and intensity, indicating a stable and reversible lithiation/delithiation response after the initial electrochemical activation of the laser-processed material [7].

The cycling performance was evaluated between 1 and 3 V at different C-rates, as shown in **Figure 12(B)**. The specific capacity progressively decreased with increasing C-rate, consistent with the behaviour reported for other TNO materials [7], due to the kinetic limitations associated with $Li^+$ insertion and extraction at higher current densities. However, when the rate was returned to 1C, the capacity recovered, suggesting that the losses observed at higher rates were primarily related to kinetic limitations rather than irreversible electrode degradation. Subsequent prolonged cycling at 1C revealed a gradual capacity decay over nearly 700 cycles, while the Coulombic efficiency remained close to 100% throughout most of the experiment. Overall, these results demonstrate that the 1 μJ material retains electrochemical activity over extended cycling and exhibits good tolerance to variations in C-rate, supporting its potential as a TNO-based anode material.

To investigate the electrode kinetics during cycling, potentiostatic electrochemical impedance spectroscopy (PEIS) measurements were performed (**Figure 12(C)**), and the resulting Nyquist plots were fitted using the equivalent circuits shown in **Figure 12(D)** (fitting parameters provided in **Table SM2**). In these circuits, R2 represents the effective series resistance, R1 the charge-transfer resistance, CPE1 the non-ideal double-layer response, and CPE2 the onset of diffusion-controlled behaviour, while the additional R3 introduced after cycling accounts for increasing transport resistance within the porous electrode. The initial plot exhibits a depressed semicircle followed by a low-frequency diffusion-related contribution. After 60 cycles, the semicircle slightly decreases, with a drop in R1, indicating electrode activation and improved interfacial kinetics. However, prolonged cycling leads to a progressive impedance increase, accompanied by a shift of the low-frequency response towards higher impedance values, indicating increasing limitations to $Li^+$ transport. This ageing trend is also reflected in the Bode plots (**Figure SM9**), showing an increased impedance modulus and a shift of the phase response towards lower frequencies. These changes correlate with the gradual capacity decay observed during long-term cycling, suggesting that electrode ageing is primarily governed by deteriorating charge-transfer kinetics and ionic transport rather than a loss of electrochemical reversibility.

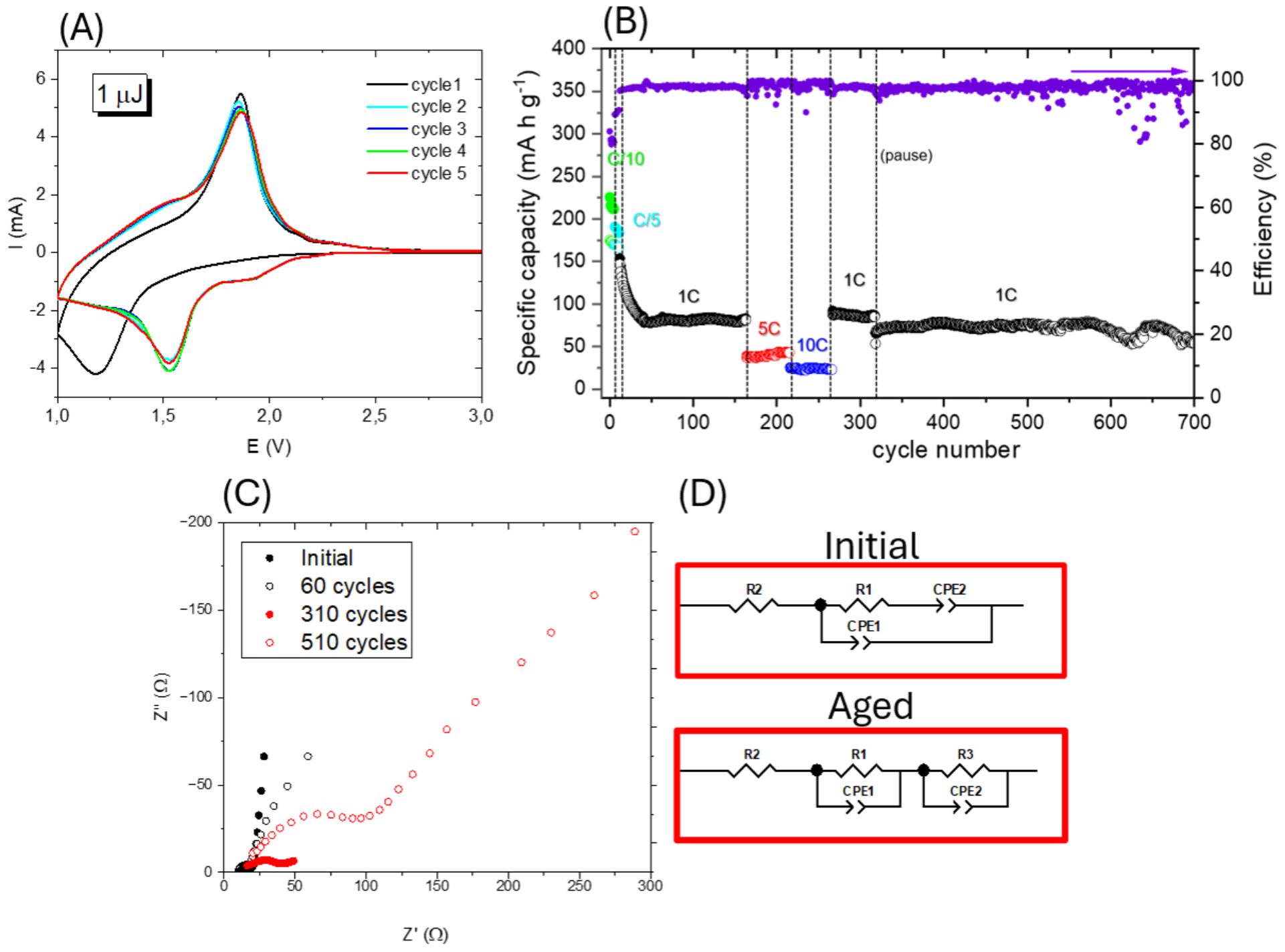

**Figure 12.** Electrochemical characterization of the non-transferred material. (A) Cyclic voltammograms recorded between 1 and 3 V at a scan rate of 0.1 mV/s for the first five cycles. (B) Rate capability and long-term galvanostatic cycling performance at different C-rates, showing the specific discharge capacity and Coulombic efficiency as a function of cycle number. (C) Nyquist plots recorded after the initial electrochemical activation and after 60, 310 and 510 galvanostatic cycles. (D) Equivalent circuits used to fit the initial impedance response (top) and the spectra obtained after cycling (bottom).

The specific capacities are lower than those reported for optimized phase-pure or nanostructured TNOs, particularly at high C-rates [39]. The reversible redox response, capacity recovery after high-rate cycling, and sustained Coulombic efficiency demonstrate that the laser-processed material retains the characteristic electrochemical functionality of TNO-based anodes. Similar kinetic and transport limitations have been mitigated in TNOs through optimization of particle size, crystallinity, defect concentration, and electronic conductivity [40]. In summary, the low-energy laser-sintered layers show a sufficient performance acting as electrode for a LIB, which might be enhanced through post-processing treatment of the layers.

## 4. Conclusions

Following an ultrafast laser energy efficient process, we have been able to produce different TNO structures directly from a non-sintered mixture of powder precursors. Three different energy regimes were identified according to the resulting material: a low-energy regime (E = 1 µJ, non-transference scenario), leading to $Ti_2Nb_{10}O_{29}$ layers together with residual $TiO_2$; an intermediate-energy regime (E = 4 µJ, transference scenario); leading to $TiNb_2O_7$ layers and Nb-enriched contribution; and a high-energy regime (E = 6 µJ, transference scenario), where the main detected crystalline phase is $TiNb_2O_7$. The high-energy regime showed a strong sensitivity to the processing conditions and to the increase in the size of the processed area, resulting in lower thickness homogeneity. In contrast, the low- and intermediate-energy regimes provided more stable processing conditions.

The results reveal a strong dependence of TNO phase formation on laser energy and thermal accumulation, progressing from incomplete precursor conversion at low energies to phase-pure $TiNb_2O_7$ under optimized high-energy conditions. The proposed formation mechanism involves rapid laser-induced melting, material transfer, and re-solidification, resembling a LIBT process.

The different materials obtained require diverse integration strategies. The low-energy material can be recovered and prepared as an electrode, whereas the intermediate- and high-energy regime directly produces continuous layers supported on the FSG. The transferred 4 µJ layers showed an improved electrical performance that overcomes the usual conductivities found in the literature. On the other hand, electrochemical measurements of the 1 µJ material confirmed reversible lithiation/delithiation behaviour and long-term electrochemical activity.

These results demonstrate the potential of fs-laser powder processing as a rapid route for the fabrication of functional TNOs, although some optimization of the structures is needed to produce competitive performance in real applications (specially for the adjusted high-energy regime). An advantage of the proposed technique is that it is not only an efficient way to quickly produce TNO films, but it can be also used as the post-processing tool that may induce annealing, polishing or other processes usually employed to enhance the electrochemical and/or electrical performance of TNOs. Besides, this process may be extended to the production of other ceramic layers from their precursor oxides.

## Acknowledgements:

This work has been supported by grant HyperSolar (TED2021-130894B-C22) funded by MICIU/AEI/10.13039/501100011033 and by the 'European Union NextGenerationEU/PRTR'; by grants ULS_PSB (PID2020-112770RB-C21) and HyperSpec (PID2023-148178OB-C22) funded by MICIU/AEI/10.13039/501100011033; by grant MATRIX-CM (TEC-2024/TEC-85) funded by Consejería de Educación, Ciencia y Universidades (Comunidad de Madrid); and by projects PR27/25-32482 and PR27/25-32452 funded by Universidad Complutense de Madrid. A. Calvo-Villoslada acknowledges the research contract under Investigo Program (CT19/23-INVM-34) of Ministerio de Trabajo y Economía Social. Authors want to acknowledge assistance from CAI de Técnicas Químicas (Unidad de Difracción de Rayos X) from Universidad Complutense de Madrid for XRD measurements. Authors want to acknowledge the help provided by Dr. Rocío Ariza, Joaquín Calbet, Gabriel Caballero and Sofía Alonso-Martínez with the scientific discussion.

# Supplementary Material File:

# Femtosecond laser-controlled production of functional titanium niobate layers from $TiO_2$/$Nb_2O_5$ powders

Gonzalo Gomez-Munoz[1*], Aarón Calvo-Villoslada[2], Inmaculada Álvarez-Serrano[3], María Luisa López[3], Paloma Fernandez[2], Javier Solis[1], Belén Sotillo[2**].

[1]Laser Processing Group (LPG), Instituto de Óptica "Daza de Valdés" (IO-CSIC), CSIC, Serrano 121, 28006, Madrid, Spain.
[2]Departamento de Física de Materiales, Facultad de Ciencias Físicas, UCM, Plaza de Ciencias 1, 28040, Madrid, Spain.
[3]Departamento de Química Inorgánica, Facultad de Ciencias Químicas, UCM, Plaza de Ciencias 2, 28040, Madrid, Spain.

[*]gonzalo.gomez@io.cfmac.csic.es
[**]bsotillo@ucm.es

Additional figures and tables are included in the current document in order to help and give a broader point of view to the reader of the main text of the paper. The descriptions of these additional data are included in the captions each figure/table and referred directly in the main text of the article as 'see Figure SM#' or 'see Supplementary Material File, Figure SM#'.

Following the strategy reported in our previous work [18], the precursor powders were compacted under a compressive load of 1 T to form disk-shaped pellets. Irradiation conditions similar to those described in our previous work were tested, i.e., a repetition rate of 100 kHz, a pulse energy of 4 μJ, and a scanning speed of 0.1 m/s. However, we observed that the conditions optimized for sintering $Nb_2O_5$ powders (see **Figure SM1(A)**) did not enable efficient incorporation of Ti into the niobium oxide matrix, as confirmed by Raman measurements (see **Figure SM1(B)**). Increasing the pulse energy or repetition rate was therefore required to promote Ti diffusion, but both parameters led to material ejection rather than the formation of a TNO layer (see **Figure SM1(C)**). In addition, the layers produced on the pellets could not be easily separated from the underlying powder, preventing their use in functional devices.

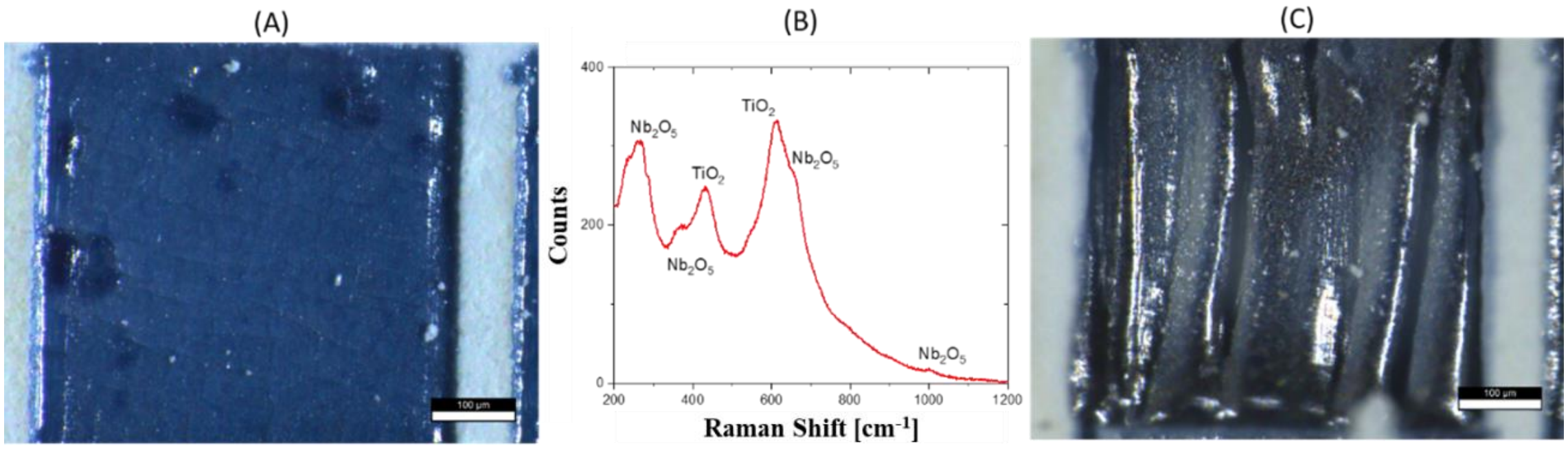

**Figure SM1.** Preliminary fs-laser processing of compacted $TiO_2$-$Nb_2O_5$ pellets. (A) Appearance of the pellet after irradiation using the conditions previously optimized for $Nb_2O_5$ processing ($E = 4$ μJ, $f_{rep} = 100$ kHz and $v = 0.10$ m/s). (B) Raman spectra of the processed region, showing limited incorporation of Ti into the niobium-oxide matrix under these conditions. (C) Representative morphology obtained after increasing the pulse energy and/or repetition rate, showing material ejection instead of continuous layer formation.

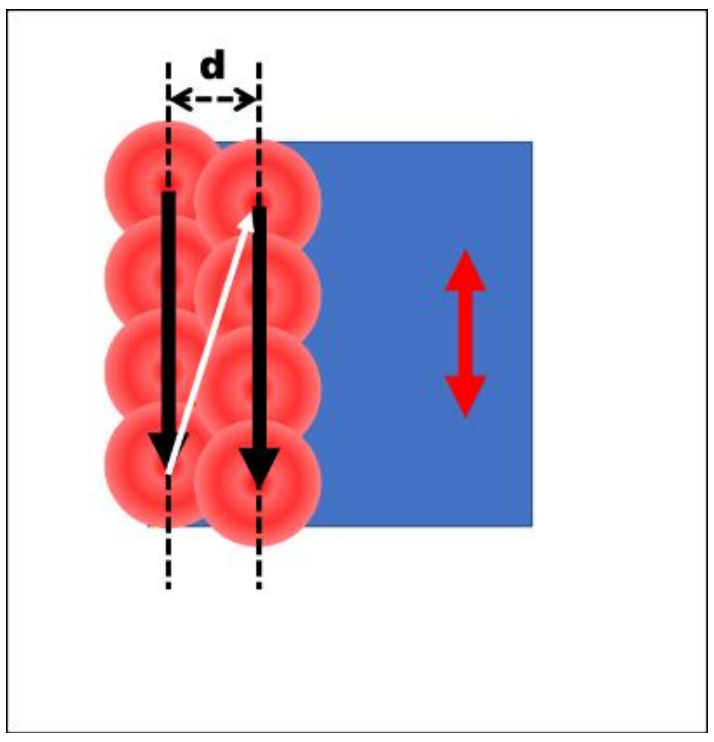

**Figure SM2.** Scheme of irradiation of one area. The white square represents the powder, and the blue one the fs-laser processed area. Each red circle denotes a laser pulse reaching the surface of the powder. The scanning direction is vertical, from the top part of the sample to the bottom part (as indicated with the black arrows), and the polarization of the light is parallel (as indicated with the red arrow). Each time a line is scanned, the laser turns off and jumps to the beginning of the next line, separated a distance d. The white arrow shows this movement of the scanner motors to precisely locate the next pulse in the desired position of the sample (upon real irradiations, there would not be a vertical shift with respect to the previous line, but here it was introduced to help with scheme).

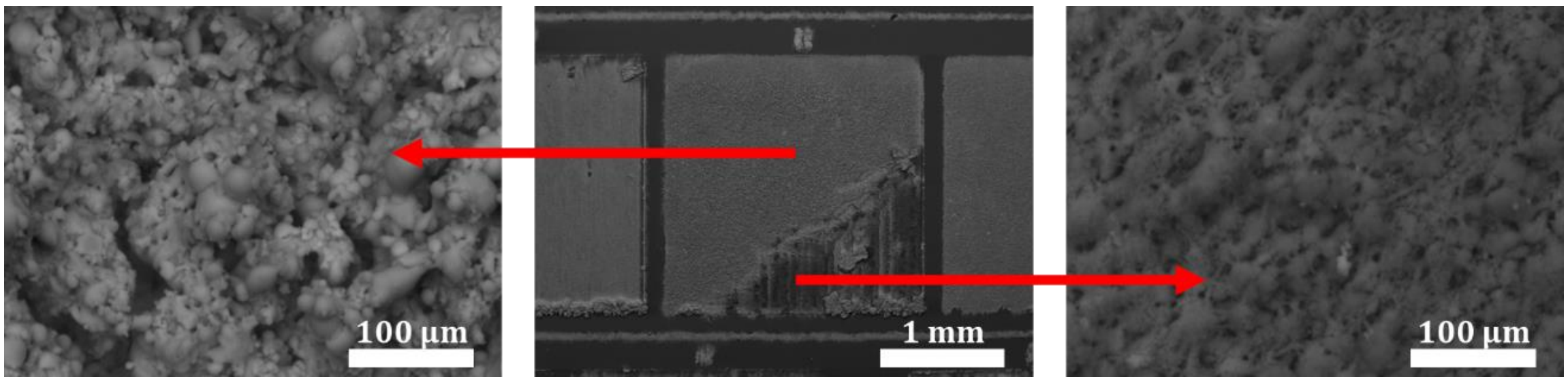

**Figure SM3.** SEM pictures of E = 5 µJ laser-processed area in the E-study. The left picture illustrates the appearance of the sintered material with a 4 µJ colour (blue if seen by OM) and the right one the appearance of the sintered material with a 6 µJ colour (white if seen by OM), and the red arrows identify which part of the 5 µJ area presents which appearance.

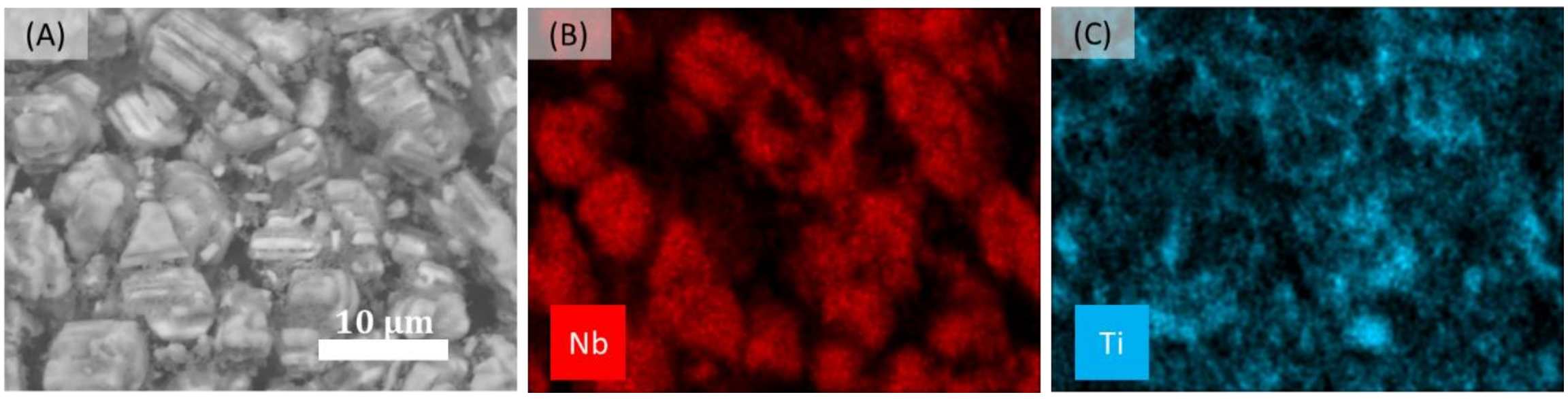

**Figure SM4.** (A) Backscattered electron SEM picture of the powder mixture as ball-milled. (B) Nb- and (C) Ti-distribution maps measured by EDX-SEM on image (A). The EDX quantification averages the following composition for the powder: Oxygen = 63.8 at.%, Niobium = 24.0 at.%, Titanium = 12.2 at.%.

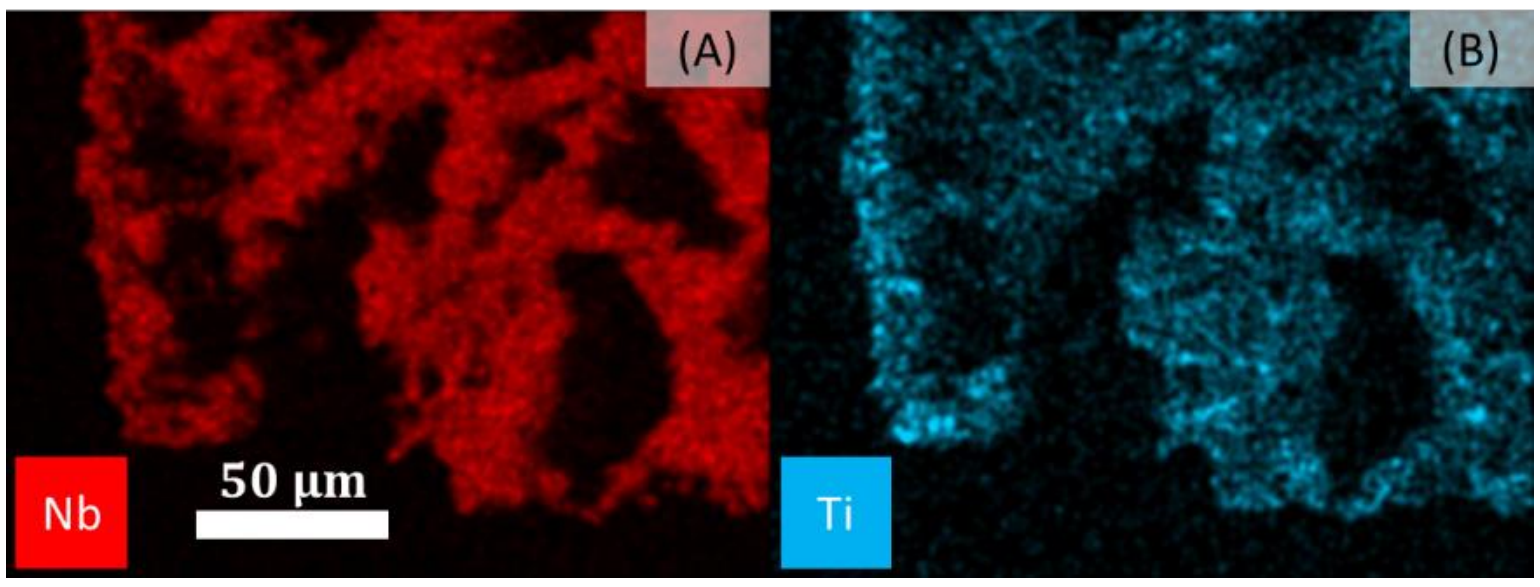

**Figure SM5.** (A) Nb- and (B) Ti-distribution maps measured by EDX-SEM on image **Figure 5** of the main text.

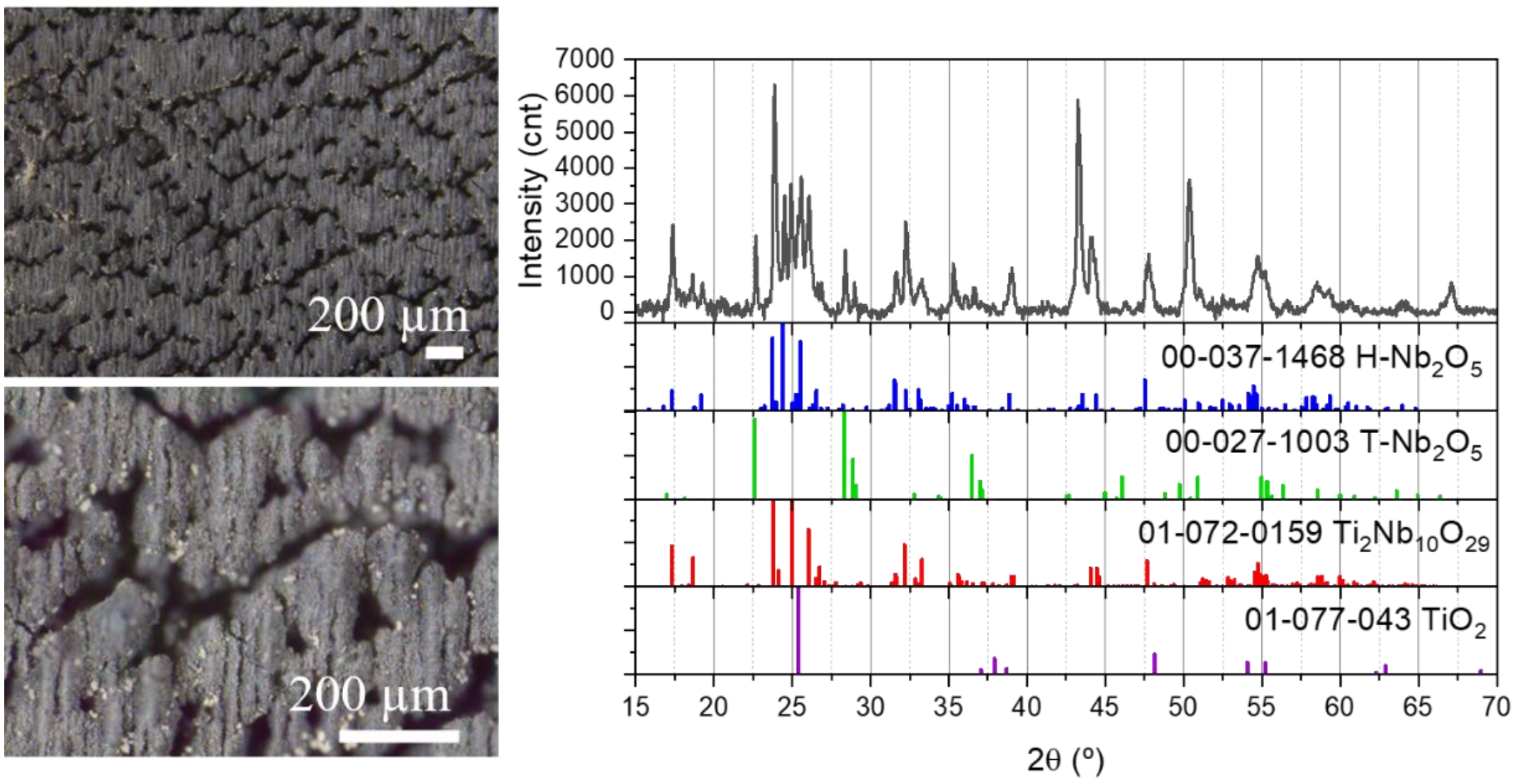

**Figure SM6**. Non-transferred fs-laser sintered material upon E = 4 µJ, v = 0.10 m/s, d = 4 µm. In the left part of the figure the powder is shown as SEM pictures (both represent the same area with more or less magnification). In the right part of the figure, the XRD measured (grey) appear compared to the XRD patterns of $H\text{-}Nb_2O_5$ (blue), $T\text{-}Nb_2O_5$ (green), monoclinic $Ti_2Nb_{10}O_{29}$ (red) and anatase $TiO_2$ (pink) extracted from the PDF references indicated on top of each pattern.

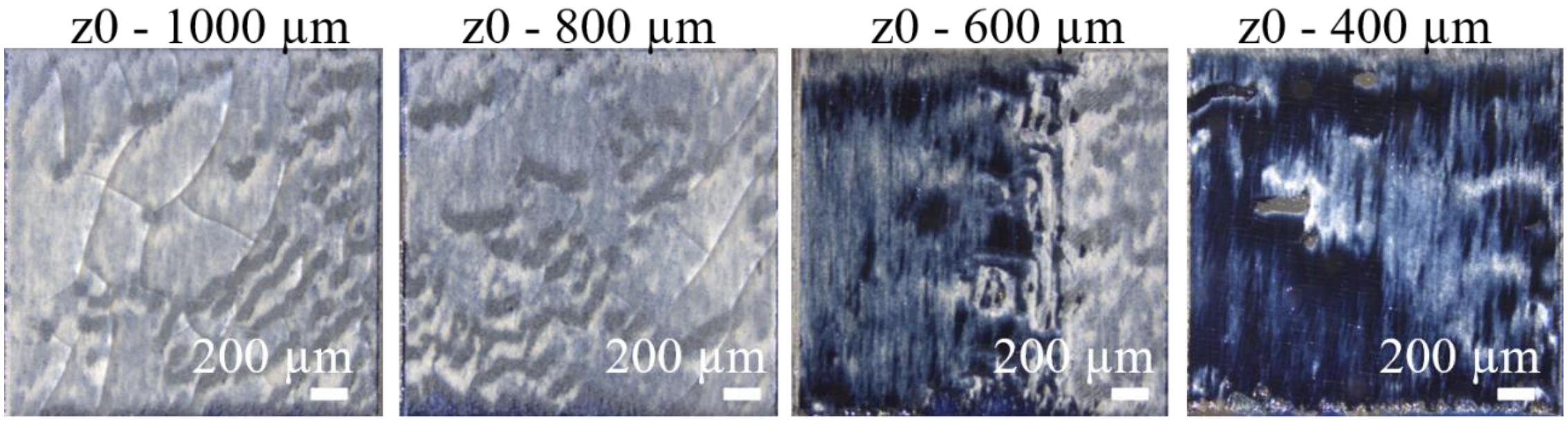

**Figure SM7.** Effect of focal position on material transference for E = 5 µJ, v = 0.10 m/s and d = 4 µm. Optical micrographs of the layers observed through the FSG. As it is patent, the colour of the processed material changes when displacing the focus inside of the powder, evidencing the importance of the focus search criterium when employing this technique.

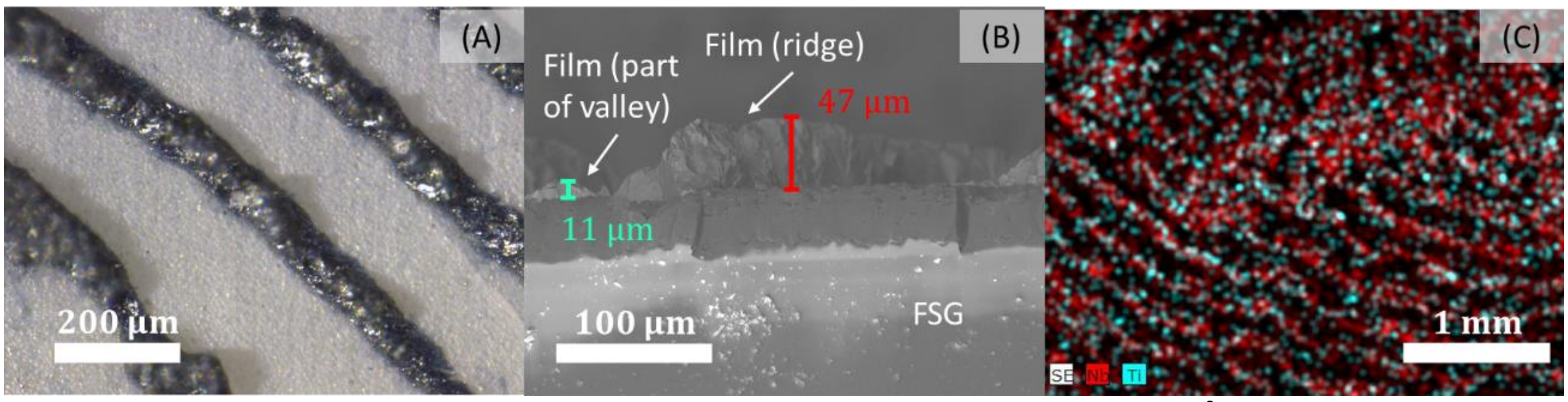

**Figure SM8.** (A) OM picture of the corrugations produced in a 6 µJ, 0.05 m/s, 6 × 6 $mm^2$ area. (B) X-BSE-SEM of the same area, showing the thickness of the layer on the ridges (red) and on the valleys (turquoise) of the corrugations. (C) Low magnification top-view EDX-SEM pictures of the previous film, with (red) Nb- and (blue) Ti-distribution maps superimposed on the BSE picture.

**Table SM1**. Mean values measured in each 100 measurements set for the sheet resistance ($R_□$) and the bulk resistivity (ρ). Each film is related to a set of 100 measurements, and each mean measured value has a related error Δ(x). The sample and the set are identified equally (i.e 'Sample n' would be 'Set n').

| Sample | $R_□$ [Ω/□·$10^5$] | Δ($R_□$) [Ω/□·$10^3$] | ρ [Ω·m] | Δ(ρ) [Ω·m·$10^{-3}$] |
|---|---|---|---|---|
| 1 | 1.183 | 0.11 | 1.01 | 0.89 |
| 2 | 1.212 | 0.19 | 1.03 | 1.65 |
| 3 | 2.990 | 0.11 | 2.54 | 0.95 |
| 4 | 4.056 | 0.34 | 3.45 | 2.92 |
| 5 | 4.888 | 0.23 | 4.15 | 1.98 |
| 6 | 2.115 | 0.16 | 1.80 | 1.37 |
| 7 | 1.879 | 0.71 | 1.60 | 6.04 |
| 8 | 1.836 | 0.10 | 1.56 | 0.86 |
| 9 | 1.593 | 0.09 | 1.35 | 0.79 |
| 10 | 2.090 | 0.15 | 1.78 | 1.31 |
| 11 | 2.153 | 0.34 | 1.83 | 2.92 |
| 12 | 4.949 | 0.42 | 4.21 | 3.56 |
| 13 | 3.602 | 2.22 | 3.06 | 18.89 |
| 14 | 3.211 | 0.20 | 2.73 | 1.70 |
| 15 | 3.067 | 0.71 | 2.61 | 6.06 |
| 16 | 0.345 | 0.06 | 0.29 | 0.48 |
| 17 | 1.281 | 0.08 | 1.09 | 0.68 |
| 18 | 1.230 | 1.31 | 1.05 | 11.13 |
| 19 | 0.515 | 0.04 | 0.44 | 0.31 |
| 20 | 0.648 | 0.13 | 0.55 | 1.08 |

**Table SM2**. Equivalent-circuit fitting parameters obtained from PEIS measurements of the 1 μJ electrode.

| **008**<br>**Equivalent circuit** | **Initial**<br>**(1)** | **60 cycles**<br>**(2)** | **310 cycles**<br>**(1)** | **510 cycles**<br>**(1)** |
|---|---|---|---|---|
| **R2 (Ω)** | 9.60 | 10.14 | 10.63 | 14.42 |
| **R1 (Ω)** | 11.19 | 7.13 | 25.87 | 66.69 |
| **CPE1**<br>**T ($Ω^{-1}·s^P$)**<br>**P** | <br>$9.23·10^{-5}$<br>0.82 | <br>$4.15·10^{-4}$<br>0.62 | <br>$1.15·10^{-4}$<br>0.59 | <br>$2.57·10^{-5}$<br>0.76 |
| **R3 (Ω)**<br>**CPE2**<br>**T ($Ω^{-1}·s^P$)**<br>**P** | <br><br>$2.7·10^{-3}$<br>0.91 | 263<br><br>$2.6·10^{-3}$<br>0.82 | <br><br>$5.1·10^{-2}$<br>0.3 | <br><br>$1.42·10^{-3}$<br>0.49 |

The initial spectrum shows a depressed semicircle followed by the onset of a low-frequency diffusion-related contribution. After 60 cycles, the semicircle becomes slightly smaller and the fitted charge-transfer resistance R1 decreases from approximately 11.2 to 7.1 Ω, indicating an initial electrochemical activation and improved interfacial kinetics. Upon prolonged cycling, however, the impedance progressively increases with the charge-transfer resistance R1 reaching approximately 26 Ω after 310 cycles and 67 Ω after 510 cycles. The low-frequency response also shifts towards higher impedance values, indicating increasing limitations to ionic transport within the porous electrode. This evolution is consistent with the gradual capacity decay observed during long-term galvanostatic cycling and suggests that electrode ageing is mainly associated with deterioration of charge-transfer kinetics and increasing transport limitations rather than with a loss of electrochemical reversibility.

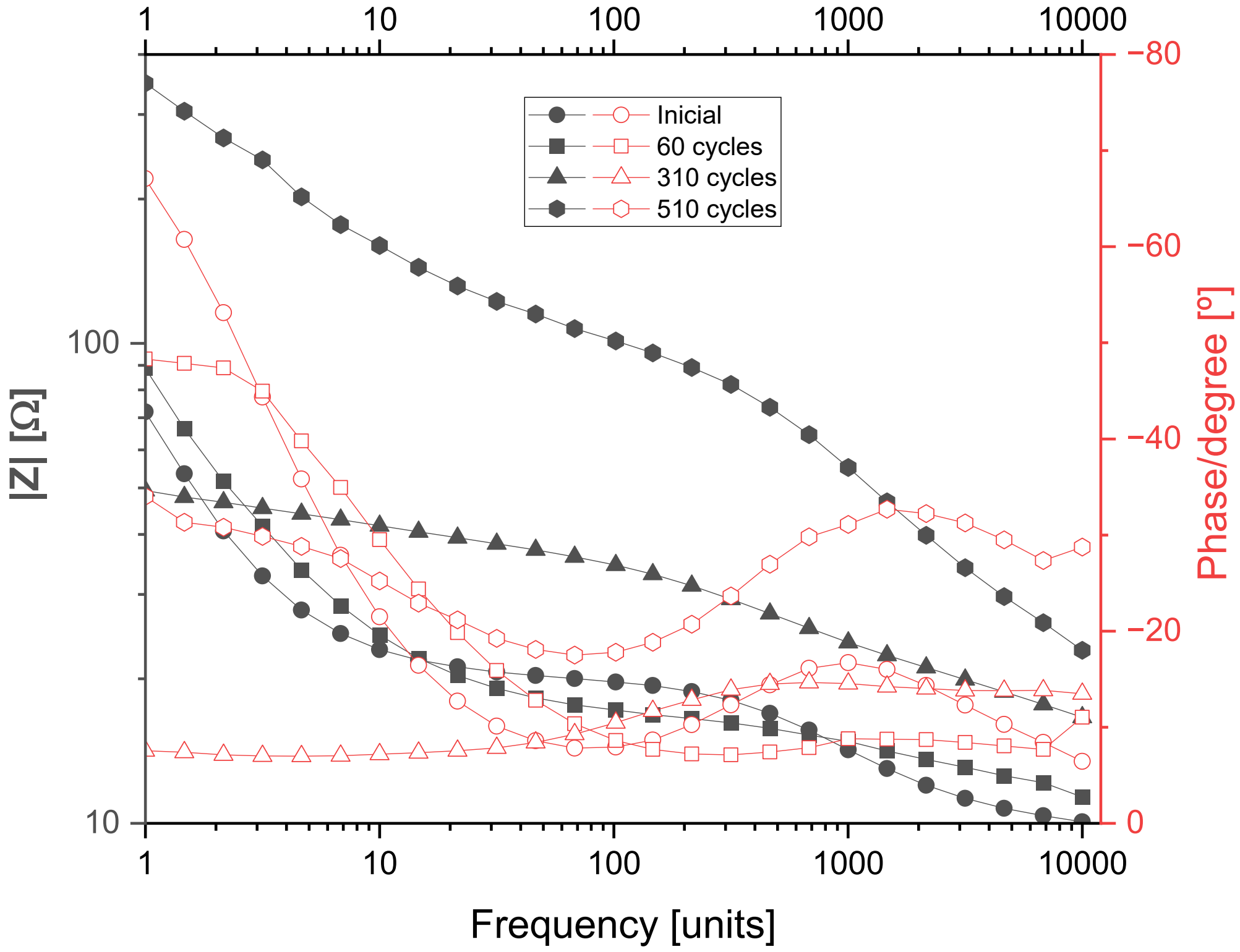

**Figure SM9**. Bode representation of the PEIS measurements for the 1 μJ electrode recorded after the initial electrochemical activation and after 60, 310 and 510 galvanostatic cycles. The impedance modulus, |Z|, and phase angle are shown as a function of frequency, illustrating the progressive increase in impedance and the shift of the characteristic relaxation response during long-term cycling.